\documentclass{article}
\usepackage[utf8]{inputenc}
\usepackage[figuresright]{rotating}
 \usepackage{bm}
\usepackage{amsmath}
\usepackage{xcolor, soul}
\usepackage{cancel}
\usepackage{lscape}
\usepackage{multirow}
\usepackage{natbib}
\usepackage{authblk}
\usepackage{listings}
\usepackage{afterpage}
\usepackage[margin=1in]{geometry}

\title{\textbf{Bayesian Estimation of Two-Part Joint Models for a Longitudinal Semicontinuous Biomarker and a Terminal Event with R-INLA: Interests for Cancer Clinical Trial Evaluation}}

\author[1,2]{Denis Rustand}
\author[2]{Janet van Niekerk}
\author[2]{H\aa vard Rue}
\author[3]{Christophe Tournigand}
\author[1]{Virginie Rondeau}
\author[4]{Laurent Briollais}
\affil[1]{Biostatistic Team, Bordeaux Population Health Center, ISPED, Centre INSERM U1219, Bordeaux, France}
\affil[2]{Statistics Program, Computer, Electrical and Mathematical Sciences and Engineering Division,\\
King Abdullah University of Science and Technology (KAUST), Thuwal 23955-6900, Kingdom of Saudi Arabia}
\affil[3]{Hopital Henri Mondor, Creteil, France}
\affil[4]{Lunenfeld-Tanenbaum Research Institute, Mount Sinai Hospital, Dalla Lana School of Public Health (Biostatistics), University of Toronto, 600 University Ave., Ontario M5G 1X5, Canada}
\date{}                     
\begin{document}

\maketitle

\begin{abstract}
Two-part joint models for a longitudinal semicontinuous biomarker and a terminal event have been recently introduced based on frequentist estimation. The biomarker distribution is decomposed into a probability of positive value and the expected value among positive values. Shared random effects can represent the association structure between the biomarker and the terminal event. The computational burden increases compared to standard joint models with a single regression model for the biomarker. In this context, the frequentist estimation implemented in the R package \textbf{frailtypack} can be challenging for complex models (i.e., large number of parameters and dimension of the random effects). As an alternative, we propose a Bayesian estimation of two-part joint models based on the Integrated Nested Laplace Approximation (INLA) algorithm to alleviate the computational burden and fit more complex models. Our simulation studies confirm that \textbf{INLA} provides accurate approximation of posterior estimates and to reduced computation time and variability of estimates compared to \textbf{frailtypack} in the situations considered. We contrast the Bayesian and frequentist approaches in the analysis of two randomized cancer clinical trials (GERCOR and PRIME studies), where \textbf{INLA} has a reduced variability for the association between the biomarker and the risk of event. Moreover, the Bayesian approach was able to characterize subgroups of patients associated with different responses to treatment in the PRIME study. Our study suggests that the Bayesian approach using \textbf{INLA} algorithm enables to fit complex joint models that might be of interest in a wide range of clinical applications.
\end{abstract} 

\section{Introduction}
\label{s:intro}
Estimation of joint models for longitudinal and time-to-event data were initially introduced using maximum likelihood estimation (\citet{wulfsohn1997joint, henderson2000joint, Song02, Chi06}). It was further developed within the Bayesian framework in situations where maximum likelihood estimation with asymptotic assumptions faces nonidentifiability issues. It allows flexible and more complex association structures and can handle multiple longitudinal outcomes (\citet{andrinopoulou2016bayesian}). Bayesian joint models can be fitted with the R package \textbf{JMbayes} (\citet{rizopoulos2016r}), which has been used in many biomedical researches (\citet{lawrence2015joint}), among other packages (e.g. \textbf{rstanarm}, \citet{muth2018user}). Bayesian estimation is usually based on MCMC techniques (\citet{hanson2011predictive, r2003bayesian, xu2001joint, Rizopoulos11}), which can have slow convergence properties. The Integrated Nested Laplace Approximation (INLA) algorithm has been recently introduced as an alternative to MCMC techniques for latent Gaussian models (LGM) (\citet{Rue09, martins2013bayesian}). Many statistical models for spatial statistics, time series, etc., can be formulated as LGMs. A key feature of \textbf{INLA} is to provide approximations of the posterior marginals needed for Bayesian inference very efficiently and that still remain very accurate compared to MCMC methods (\cite{Rue17}). By formulating complex joint models as LGMs, \textbf{INLA} can be used to fit these models as developed recently (\citet*{vanNiekerk19a, vanNiekerk19b}). For the two-part joint model, \textbf{INLA} is yet to be used for inference.\\

Two-part joint models (TPJMs) for a longitudinal semicontinuous biomarker and a terminal event have been recently introduced (\citet{Rustand20}) and applied to the joint analysis of survival times and repeated measurements of the Sum of the Longest Diameter of target lesions (SLD), which is a biomarker representative of tumor burden in cancer clinical trials. The TPJM uses a conditional two-part joint model that decomposes the biomarker distribution into a binary outcome (zero vs. positive value) fitted with a logistic mixed effects model and a continuous outcome (positive values only) fitted with either a linear mixed effects model on the log-transformed outcome or a lognormal mixed effects model (\citet{Rustand21}). The ``conditional'' form of the two-part model gives the effect of covariates on the mean biomarker value conditional on a positive value in the continuous part. An alternative marginal model has recently been proposed to get the effect of covariates on the (unconditional) mean of the biomarker. A drawback of the marginal two-part model is that it may lead to arbitrary heterogeneity and provide less interpretable estimates on the conditional mean of the biomarker among positive values (\citet{Smith14}). In this article, we focus on the conditional two-part joint model, simply referred to as TPJM in what follows. The association with the survival model can be specified in terms of shared random effects, i.e., random effects that are shared between the different components of the models including the binary and continuous parts of the model and the survival component. The TPJM is particularly interesting for cancer clinical trials evaluation because it can help characterizing subgroups of patients who can benefit from a specific treatment (i.e., patients who had a complete tumor removal and no regrowth of the tumor would have a zero value of the biomarker and this is captured by the binary part of the model) and subgroups that do not respond or disbenefit from treatment, as it is sometimes observed when patients develop a resistance to chemotherapies (this will be captured by the continuous part of the model). These features cannot be modeled adequately by alternative strategies that assume a unique distribution for the zeros and the positive tumors size values (e.g., Tobit or Tweedie model, see \citet{kurz2017tweedie}). Besides, the effect of treatment on patients' subgroups can be missed and may lead to biased conclusions as illustrated in \citet{Rustand20}.

An important limitation of the TPJM is the estimation procedure that requires a numerical approximation of the random effects distribution, which can lead to long computation times and convergence issues with high-dimensional parameter settings and complex association structures between the different components of the TPJMs. In this article, we propose an efficient Bayesian estimation procedure for the TPJM which relies on INLA algorithm, as implemented in the R package \textbf{INLA}. In practice, we used the R package \textbf{INLAjoint} that facilitates the fit of joint models with \textbf{INLA} and make it more user-friendly, as \textbf{INLA} was not developed specifically for this class of models. The Bayesian inference is compared to the frequentist estimation of the TPJM available in the R package \textbf{frailtypack} (\citet{Krol17}). The remainder of the article is structured as follows: in Section 2, we describe the TPJM and introduce the frequentist and Bayesian estimations. In Section 3, we present a simulation study to assess the performance of these two estimation strategies in terms of bias, coverage probability, computation time and convergence rate. An application to two randomized clinical trials each comparing two treatment strategies in patients with metastatic colorectal cancer is proposed in Section 4 and we conclude with a discussion in Section 5.

\section{Estimation of the conditional two-part joint model}
\label{sec:tpjm}

\subsection{Model specification}
\label{sec:ctpjm}
Let $Y_{ij}$ denote the biomarker measurement for individual $i (i=1, ..., n)$ at visit $j (j=1, ..., n_i)$, $T_i$ denotes the survival time and $\delta_i$ the censoring indicator for individual $i$. We use a logistic mixed effects model for the probability of a positive value of the biomarker and a linear mixed effects model for the conditional expected biomarker value. A proportional hazards survival model specifies the effect of covariates on survival time, adjusted for the individual heterogeneity captured in the biomarker model. The complete model is defined as follows: 
\begin{equation*}
\left\{  \begin{array}{lc}
   \eta_{Bij}=\text{Logit}(\text{Prob}(Y_{ij}>0))=\bm{X}_{Bij}^\top \bm{\alpha} + \bm{Z}_{Bij}^\top \bm{a}_i& \textrm{(Binary part),}\\
 \eta_{Cij}= \text{E}[\log(Y_{ij})|Y_{ij}>0]=\bm{X}_{Cij}^\top \bm{\beta} +  \bm{Z}_{Cij}^\top \bm{b}_i & \textrm{(Continuous part),}\\
  \lambda_i(t)=\lambda_{0}(t)\ \exp(\eta_{Si})=\lambda_{0}(t)\ \exp\left(\bm{X}_i^\top\ \bm{\gamma} + \bm{a}_i^\top \bm{\varphi}_a+\bm{b}_i^\top \bm{\varphi}_b \right)& \textrm{(Survival part),}\\
   \end{array}
\right. 
 \label{eq3}
 \end{equation*}

\noindent
where $\bm{X}_{Bij}$, $\bm{X}_{Cij}$ and $\bm{X}_i$ are vectors of covariates associated to the fixed effects $\bm{\alpha}$, $\bm{\beta}$ and $\bm{\gamma}$, respectively. Similarly, $\bm{Z}_{Bij}$ and $\bm{Z}_{Cij}$ are vectors of covariates associated to the random effects $\bm{a}_i$ and $\bm{b}_i$ in the binary and continuous parts. The random effects are shared in the survival model, with association parameters $\bm \varphi_a$ and $\bm \varphi_b$, respectively. These two vectors of random effects follow a multivariate normal distribution $\mathcal{N}(0,\bm Q_{ab}^{-1})$, with covariance matrix $\bm Q_{ab}^{-1}$, defined as

$$
\bm Q_{ab}^{-1}=
\begin{bmatrix}
\bm{\Sigma}_{aa} & \bm{\Sigma}_{ab} \\
\bm{\Sigma}_{ab} & \bm{\Sigma}_{bb}
\end{bmatrix}.
$$

They account for both the association between the three components of the model and the correlation between the repeated measurements in the longitudinal process (observations are independent conditional on the random effects). We use a log-transformation of the biomarker to account for the positivity constraint and right-skewness in the continuous part of the model. The joint distribution assumes that the vectors of random effects underlies both the longitudinal and survival process, the joint distribution of the observed outcomes for individual $i$ is defined by
\begin{align*}
p(T_i, \delta_i, \bm Y_{i} | \bm{a}_i, \bm{b}_i;\bm \Theta) &= p(T_i, \delta_i | \bm{a}_i, \bm{b}_i; \bm \Theta) \prod_{j=1}^{n_i} p(Y_{ij} | \bm{a}_i, \bm{b}_i; \bm \Theta)\\
&= p(T_i, \delta_i | \bm{a}_i, \bm{b}_i; \bm \Theta) \prod_{j=1}^{n_i} p(Y_{ij} | Y_{ij}>0; \bm{a}_i, \bm{b}_i; \bm \Theta) \ p(Y_{ij}>0; \bm{a}_i, \bm{b}_i; \bm \Theta)
 \label{jdist}
 \end{align*}
with $\bm \Theta$ the full parameter vector, including the parameters for the binary, continuous and survival outcomes, the baseline hazard function and the random effects covariance matrix, such that the full conditional distribution is given by $$p(\bm T, \bm \delta, \bm Y | \bm a, \bm b; \bm \Theta) = \prod_{i=1}^{n}p(T_i, \delta_i, \bm Y_{i} | \bm{a}_i, \bm{b}_i;\bm \Theta) .$$ The likelihood contribution for the $i$th subject can be formulated as follows
\begin{align*}
L_i(\bm \Theta | \bm Y_{i}, T_i, \delta_i)&=  \int_{\bm{a}_i} \int_{\bm{b}_i}  \prod_{j=1}^{n_i} \exp\left(\bm{X}_{Bij}^\top \bm{\alpha} + \bm{Z}_{Bij}^\top \bm{a}_i  \right)^{U_{ij}} \left(1-\frac{\exp(\bm{X}_{Bij}^\top \bm{\alpha} + \bm{Z}_{Bij}^\top \bm{a}_i )}{1+\exp(\bm{X}_{Bij}^\top \bm{\alpha} + \bm{Z}_{Bij}^\top \bm{a}_i )}\right) \\
& \times {\left\{\frac{1}{\sqrt{2 \pi \sigma^2_\epsilon}} \exp \left( - \frac{(\log(Y_{ij})-\bm{X}_{Cij}^\top \bm{\beta} -  \bm{Z}_{Cij}^\top \bm{b}_i )^2}{2 \sigma^2_\epsilon} \right)\right\}^{U_{ij}}}\\
& \times \lambda_i(T_i|  \bm{a}_i, \bm{b}_i)^{\delta_i} \exp\left( - \int_{0}^{T_i} \lambda_i(t|  \bm{a}_i, \bm{b}_i) \mathrm{d}t \right) p(\bm{a}_i, \bm{b}_i) \mathrm{d}\bm{b}_i \mathrm{d}\bm{a}_i,
   \end{align*}
where $U_{ij}=I[Y_{ij}>0]$, $\delta_i=I[T_i \text{ is not censored}]$ and $\lambda_i(t| \bm{a}_i, \bm{b}_i)=\lambda_{0}(t|  \bm{a}_i, \bm{b}_i)\ \exp \{ {\bm{X}_{i}(t)}^\top \bm{\gamma} + \bm{a}_i^\top \bm{\varphi}_a+\bm{b}_i^\top \bm{\varphi}_b \}$. \\

\subsection{Bayesian estimation of the TPJM}
\label{sec:baye}
Define $\boldsymbol{D} \equiv \{T_i, \delta_i, Y_{ij}: i=1, \cdots, n; j=1,\cdots, n_i \}$ the observation variables. The goal of the Bayesian inference is to estimate the posterior distribution $\pi (\boldsymbol{\Theta}|\boldsymbol{D})$. The joint posterior distribution $\pi(\bm \Theta | \bm D)$ is given by Bayes theorem as $$\pi(\bm \Theta | \bm D) = \frac{p(\bm D| \bm \Theta) \pi(\bm \Theta)}{\pi(\bm D)} \propto p(\bm D | \bm \Theta) \pi(\bm \Theta),$$
where $p(\bm D | \bm \Theta)$ is the likelihood and $\pi(\bm \Theta)$ is the joint prior.
The marginal likelihood $\pi(\bm D)=\int_{\bm \Theta} p(\bm D | \bm \Theta) \pi(\bm \Theta) \mathrm{d}\bm \Theta$ acts as a normalizing constant. 
The posterior marginal distribution of each parameter is then obtained by integrating out the other parameters of the model. 
In many cases, the posterior distribution is not analytically tractable and sampling-based methods like MCMC can be used. Approximate methods like INLA, provide exact approximations to the posterior at lower cost than sampling-based methods. The INLA methodology is based on the assumption that the statistical model is a latent Gaussian model, which we show in the next section for the TPJM.

\subsection{Formulation of the TPJM as a latent Gaussian model (Gaussian priors)} 
We now decompose $\bm \Theta$ into the Gaussian latent field $\bm u \equiv (\bm\eta_{B}, \bm\eta_{C}, \bm\eta_{S}, \bm a, \bm b, \bm\alpha,\bm\beta, \bm\gamma, \bm\lambda,\bm\varphi)$ and the set of hyperparameters $\bm \theta \equiv ({\bm \theta_1},{\bm \theta_2})$. The hyperparameters $\bm \theta_1$ pertain to the latent field precision structure while $\bm \theta_2$ contains the likelihood hyperparameters. The hyperparameters can follow any distribution and do not need to be Gaussian. Note that the first $\sum_{i=1}^n n_i + \sum_{i=1}^n n_i +n$ elements of $\bm u$ are the linear predictors of the TPJM and the rest of the elements are the latent unobserved variables. The linear predictors are included in the latent field so that each observation depends on the latent Gaussian field $\bm{u}$ only through one single element, which greatly simplifies the computations needed in the INLA algorithm, see \citet{Rue09, Rue17}.
We assume $\bm a_i, \bm b_i | \bm Q_{ab} \sim \mathcal{N}(0, \bm Q_{ab}^{-1})$, where $\bm Q_{ab}$ is the precision matrix of the shared random effects with corresponding hyperparameters $\bm\tau_{aa}$, $\bm\tau_{ab}$, $\bm\tau_{bb}$. We also assume $\bm \alpha \sim \mathcal{N}(0,\tau_\alpha \bm I)$, $\bm \beta \sim \mathcal{N}(0,\tau_\beta \bm I)$, $\bm \gamma \sim \mathcal{N}(0,\tau_\gamma \bm I)$ and $\bm \varphi \sim \mathcal{N}(0,\tau_\varphi \bm I)$. Let $\bm\lambda$ denote a vector of coefficients associated with a random walk order one or order two used to approximate the log-baseline hazard function $\log\left(\lambda_0(t)\right)$ of the survival model. These models are stochastic spline models with precision parameter $\tau_\lambda$.
Thus, the latent field $\boldsymbol{u}$ is multivariate Gaussian with zero mean and precision matrix $\boldsymbol{Q}(\bm \theta_1)$, i.e.,
$$\boldsymbol{u}|\bm \theta_1 \sim \mathcal{N}(0,\boldsymbol{Q}^{-1}({\bm \theta_1})).$$ Note that $\boldsymbol{Q}(\bm \theta_1)$ is a sparse matrix indexed by a low dimension of parameters $\bm \theta_1$. This implies that the latent field $\bm u$ is a Gaussian Markov random field (GMRF). The distribution of the observation variables $\boldsymbol{D}$ is denoted by $p(\bm D | \boldsymbol{u}, \bm \theta)$ and they are conditionally independent given the Gaussian random field $\bm u$ and hyperparameters $\bm \theta$ i.e.,
$$ \boldsymbol{D}|\boldsymbol{u},\boldsymbol{\theta} \sim \prod_{i=1}^n p(\bm{d_i}|u_i,\boldsymbol{\theta}).$$

\noindent Then the posterior of $(\boldsymbol{u},\boldsymbol{\theta})$ can be written as
$$ \pi(\boldsymbol{u},\boldsymbol{\theta}|\boldsymbol{D}) \propto \pi(\boldsymbol{\theta})\pi(\boldsymbol{u}|\boldsymbol{\theta})\prod_{i=1}^n p(\bm d_i|u_i,\boldsymbol{\theta}),$$ 
$$ \propto \pi(\boldsymbol{\theta}) |\boldsymbol{Q}(\bm \theta_1)|^{n/2} \exp \Big[ \frac{1}{2} \bm u^\top \boldsymbol{Q}(\bm \theta_1) \bm u + \sum_{i=1}^n \log\{ p(\bm d_i|u_i,\boldsymbol{\theta}) \} \Big].$$
This construction then shows that the TPJM is in fact an LGM since the latent field is a Gaussian Markov field and each data contribution depends on only one element of the latent field.

The main aim of INLA is then to approximate the posterior marginals $\pi(u_i|\bm D)$, $i=1,\cdots,n$, $\pi(\bm \theta|\bm D)$ and $\pi(\theta_{m} |\bm D), m=1,\cdots,dim(\bm \theta)$, as presented in the next section.

\subsection{INLA}
	The INLA methodology introduced by \citet{Rue05} is a major contribution to achieving efficient Bayesian inference, especially for complex or large models. INLA uses a unique 
	combination of Laplace Approximations and conditional distributions to approximate the joint posterior density as well as the marginals of the latent field and hyperparameters. It is thus not a sampling based method like MCMC and such.\\
	For the sake of brevity, the INLA methodology can be presented in the following three steps:
	\begin{enumerate}
		\item Approximate the marginal posterior distribution of hyperparameters using the Laplace approximation.
		$$
		\pi(\bm\theta|\pmb{D})=\frac{\pi(\pmb{u},\bm\theta|\pmb{D})}{\pi(\pmb{u}|\bm\theta,\pmb{D})}\approx \frac{\pi(\bm\theta)\pi
			(\pmb{u}|\bm\theta)p(\pmb{D}|\pmb{u},\bm\theta)}{\tilde{\pi}_G(\pmb{u}|\bm\theta,\pmb{D})}|_{u=u^*(\bm\theta)},
		$$ where ${\tilde{\pi}_G(\pmb{u}|\bm\theta,\pmb{D})}$ is the Gaussian approximation of ${{\pi}(\pmb{u}|\bm\theta,\pmb{D})}$ at the mode $u^*(\bm\theta)$ of the latent field for a given configuration of $\bm\theta$. The marginal posterior $\pi(\theta_m|\pmb{D})$ can be approximated by integrating $\bm\theta_{-m}$ out in the previous approximation (while a good approximation of $\pi(u_i|\pmb{\theta},\pmb{D})$ is required to approximate the posterior marginal $\pi(u_i|\pmb{D})$).
		\item Approximate the conditional posterior distributions of the latent field.
		$$
		\pi(u_i|\pmb{\theta},\pmb{D}) \propto \frac{\pi(\pmb{u},\bm\theta|\pmb{D})}{\pi(\pmb{u}_{-i}|u_i,\bm\theta,\pmb{D})},
		$$
		using a Gaussian approximation (option 1), or a Laplace approximation in a similar way as mentioned in step 1 (option 2) or using a ``Simplified Laplace approximation'' (\citet{Rue09}), which corrects the Gaussian approximation for location and skewness by expanding the numerator and denominator up to a third order Taylor series expansion in the Laplace approximation (option 3).
		\item Use numerical integration to approximate the marginal posterior distributions of the latent field.
		$$
		\pi(u_i|\pmb{D})\approx \sum_{h=1}^{H}\tilde{\pi}(u_i|\bm\theta_h^*,\pmb{D})\tilde{\pi}(\bm\theta^*_h|\pmb{D})\Delta_h,
		$$
		from steps 1 and 2. The integration points $\{\bm\theta^*_1, ..., \bm\theta^*_H\}$ are selected from a rotation using polar coordinates and based on the density at these points, and $\Delta_h$ are the corresponding weights. The approximation of the posterior marginal for each element of the latent field and each hyperparameter, using numerical integration, forms the ``integrated'' part of INLA algorithm while the first two steps above correspond to the ``nested Laplace'' approximation steps of INLA.
	\end{enumerate}

	\subsection{Priors for the hyperparameters, $\bm \theta$}	\label{subsec:pcprior}
	From the formulation of the TPJM as an LGM, the prior for the hyperparameters, $\pi(\pmb{\theta})$, should be specified. This prior can assume any form while keeping the TPJM an LGM. Amidst the debate about priors, \citet{Simpson17} proposed a framework to construct principled priors for hyperparameters, namely penalizing complexity (PC) priors. These priors are derived based on the distance from a complex model to a simpler (base) model, with a user-defined parameter that informs the strength of contraction towards the simpler model. This parameter defines whether the PC priors are vague, weakly informative, or strongly informative based on the departure from the base model measured by the Kullback-Leibler distance. It is based on the principle of parsimony, simplifying the interpretation of the results by ensuring that the priors do not overfit. For example, the PC prior for the precision of the random walk model (baseline hazard) is derived to contract towards infinity, which in turn results in zero variance. The base model thus is a constant level model (constant baseline hazard, point mass at zero) while the complex model stems from finite precision as a random spline with zero mean.\\
	
In our case we have various precision hyperparameters, $\{\bm\tau_{aa},\bm \tau_{ab}, \bm\tau_{bb}, \tau_\alpha, \tau_\beta, \tau_\gamma, \tau_\varphi, \tau_\lambda\}$. We assign weakly informative priors to the fixed effects such that $\tau_\alpha=\tau_\beta=\tau_\gamma=\tau_\varphi = 10^{-3}$. We thus need to formulate priors for the elements of $\bm Q_{ab}$ and $\tau_\lambda$. For all these hyperparameters (precision and correlation parameters), we assume the respective PC priors as given in \citet{Simpson17}.\\ As illustration we give the details for the precision of the random intercept in the binary part of the TPJM (i.e., $a$) assuming the model proposed in the next Section, $\tau_{a}=\Sigma_{aa}^{-1}$. The PC prior is derived as $$\pi(\tau_{a})=\frac{\rho}{2}\tau_{a}^{-3/2}\exp(-\rho\tau_{a}^{-1/2}),$$ with the user-defined scaling parameter $\rho=-\frac{\ln(v)}{w}$. This parameter is chosen based on the desired tail behaviour (or strength of contraction towards the base model $\sigma_{a}=\tau_{a}^{-1/2}=0)$ in the sense that $v$ and $w$ are such that $$P[\sigma_{a}>w]=v, \quad w>0, 0<v<1.$$ Larger values of $v$ and $w$ results in higher prior density away from the base model, whereas smaller values of $v$ places more density closer to the base model. The same principle is used for specifying $\pi(\tau_\lambda)$.

\section{Simulation study}
\label{sec:sim}

\subsection{Settings}
We designed simulation studies to compare the performances of \textbf{INLA} and \textbf{frailtypack} in terms of bias of the parameter estimates, coverage probabilities, computation time and convergence rates. The main factor driving the performance is the model complexity defined by the number of parameters. In particular, the number of correlated random effects defines the dimension of the integration that needs to be numerically approximated.  We propose four simulation scenarios where the parameter values of the simulation models and proportion of zeros are based on the results obtained from the real data analyses. The first scenario includes a random intercept in the binary and continuous parts of the TPJM that are correlated. The second simulation scenario includes an additional random effect for the individual deviation from the mean slope in the continuous part, thus 3 correlated random effects. These two first simulation scenarios include 200 individuals in each dataset, corresponding to a small sample size commonly seen in randomized clinical trials while the third simulation scenario includes 500 individuals in each dataset. The last simulation scenario includes a sample size of 200 individuals and natural cubic splines for the biomarker specified with the \textit{ns()} function in R, with a knot at the sample median of observed times. This generates 3 bases for the continuous part (1 intercept and 2 slopes separated by the knot) and 1 basis for the binary part (intercept). The bases are then assumed correlated random effects in the model. For each scenario, we generate 1000 datasets, we first sample the positive longitudinal biomarker repeated measurements from a Gaussian distribution and include the zero values sampled from a binomial distribution. The association between the probability of zero value and the positive values is given by the correlated random effects. Survival times for the terminal event are generated from an exponential baseline hazard function with a scale of $0.2$, an administrative censoring is assumed to occur at the end of the follow-up (4 years). The rate of zeros is $8\%$ (SD=$1\%$) for the first three scenarios and $5\%$ for the last scenario, which is in between what we observed in our two real datasets ($12\%$ of zeros in application 1 and $4\%$ in application 2). A zero value observation corresponds to a patient who experienced a complete disappearance of his/her target lesions and thus is extremely informative about treatment effect. The model for data generation is given by

\begin{equation*}
\left\{   \begin{array}{lc}
\text{Logit}[\text{Prob}(Y_{ij}>0)]=\alpha_0+a_{i}+\alpha_{1} \cdot time_j + \alpha_{2} \cdot trt_i + \alpha_{3} \cdot time_j \cdot trt_i,\\
  \textrm{E}[\log(Y_{ij})|Y_{ij}>0]=\beta_0 + b_{0i}+ (\beta_{1} + b_{1i}) \cdot time_j + \beta_{2} \cdot trt_{i}+ \beta_3 \cdot time_j \cdot trt_{i} + \varepsilon_{ij},\\
  \lambda_i(t|Y_{ij})=\lambda_{0}(t)\ \textrm{exp}\left(\gamma \cdot trt_i +\varphi_a \cdot a_i + \varphi_{b_0} \cdot b_{0i} +\varphi_{b_1} \cdot b_{1i}]\right),\\
   \end{array}
   \right.
\end{equation*}

   \begin{equation*}
\begin{bmatrix}
a_i \\
b_{0i} \\
b_{1i}
\end{bmatrix}
\sim MVN \left( \begin{bmatrix}
\bm 0 \\
\bm 0
\end{bmatrix} , 
\begin{bmatrix}
\bm\Sigma_{aa} & \bm\Sigma_{ab}\\
 \bm\Sigma_{ab} & \bm\Sigma_{bb}\\
\end{bmatrix}\right)
\equiv MVN \left( \begin{bmatrix}
0 \\
0 \\
0
\end{bmatrix} , 
\begin{bmatrix}
\sigma^2_a & \sigma_{a b_0}&  \sigma_{a b_1}\\
\sigma_{a b_0} & \sigma_{b_0}^2 & \sigma_{b_0 b_1}\\
\sigma_{a b_1} & \sigma_{b_0 b_1} & \sigma_{b_1}^2
\end{bmatrix}\right).
\label{eq9}
\end{equation*}

Note that in the first simulation scenario, we consider only a random intercept in the continuous part and the covariance matrix of the random effects is only defined by its first 2 lines and columns while in the second and third scenario we use the full matrix. For the last scenario, the evolution of the biomarker over time in the continuous part is captured by two splines bases, each one has an interaction with treatment and is associated to a random effect.
The baseline hazard function in the survival part of the model is approximated by a random walk model with \textbf{INLA} (\citet{martino2011approximate}) such that for $m$ bins on the time axis, $$\lambda_k - \lambda_{k-1}\sim N(0, \tau_{\lambda}),$$
where the PC prior (see Section \ref{subsec:pcprior}) is used is used for the prior of $\tau_{\lambda}$.\\

The random walk order one model is a stochastic smoothing spline that smooths based on first order differences. The number of bins are not influential (as opposed to knots of other splines) since an increase in the number of bins only results in an estimate closer to the stochastic model. In the simulations and applications, we use the random walk order two model that provides a smoother spline since the smoothing is then done on the second order.
See \citet{vanSankhya} for more details on the use of these random walk models as Bayesian smoothing splines.
This approximation is different with \textbf{frailtypack} that uses cubic M-splines with $5$ knots. A penalization ensures that the baseline hazard is smooth (a smoothing parameter is chosen using an approximate cross-validation criterion from a separate Cox model). The Levenberg-Marquardt algorithm, a robust Newton-like algorithm maximizes the log-likelihood function with \textbf{frailtypack} (\citet{Marquardt63}). The convergence of the algorithm depends on three conditions: The difference between the log-likelihood, the estimated coefficients and the gradient of the log-likelihood of two consecutive iterations must be under $10^{-3}$. These convergence criteria avoids spurious convergence, making this algorithm more reliable than classical alternatives (e.g., EM or BFGS, see \citet{philipps2021robust}). We use a Monte Carlo approximation for the approximation of the integrals over the random effects in the likelihood function, with 1000 integration points which is a reasonable tradeoff between the precision of the approximation and computation time (using 2000 integration points doubled the computation time with negligible improvements of the results). The simulation studies are performed with 80 CPUs, \textbf{frailtypack} uses Message Passing Interface (MPI) for parallel computation while the conjunction of \textbf{INLA} with the \textbf{PARDISO} library provides a high performance computing environment with parallel computing support using OpenMP (\citet{Schenk04}). In practice, the 80 CPUs are mainly useful to reduce the computation time with \textbf{frailtypack} because the computation time with \textbf{INLA} is very low regardless of the number of threads because of the small sample size and number of hyperparameters.

\subsection{Results}
In the results, we are comparing a Bayesian and a frequentist method, which each has usually its own evaluation criteria. Frequentist bias is used to evaluate the results from \textbf{frailtypack} while the plausibility of the result based on $95\%$ credible intervals is used to evaluate the results from \textbf{INLA} (\citet{hespanhol2019understanding}). However, we are interested in the Bayesian approximation of the MLE (i.e. non informative priors) and therefore we provide an interpretation in this context (\citet{walker1969asymptotic}). These 95\% credible intervals are obtained by truncating the tails of the posterior distributions of each parameter with \textbf{INLA} while 95\% confidence with \textbf{frailtypack} are obtained with the inverse Hessian of the model, assuming Gaussian distribution for each parameter.

\begin{figure}[!ht]
\centering
\includegraphics[width=17cm, height=8cm]{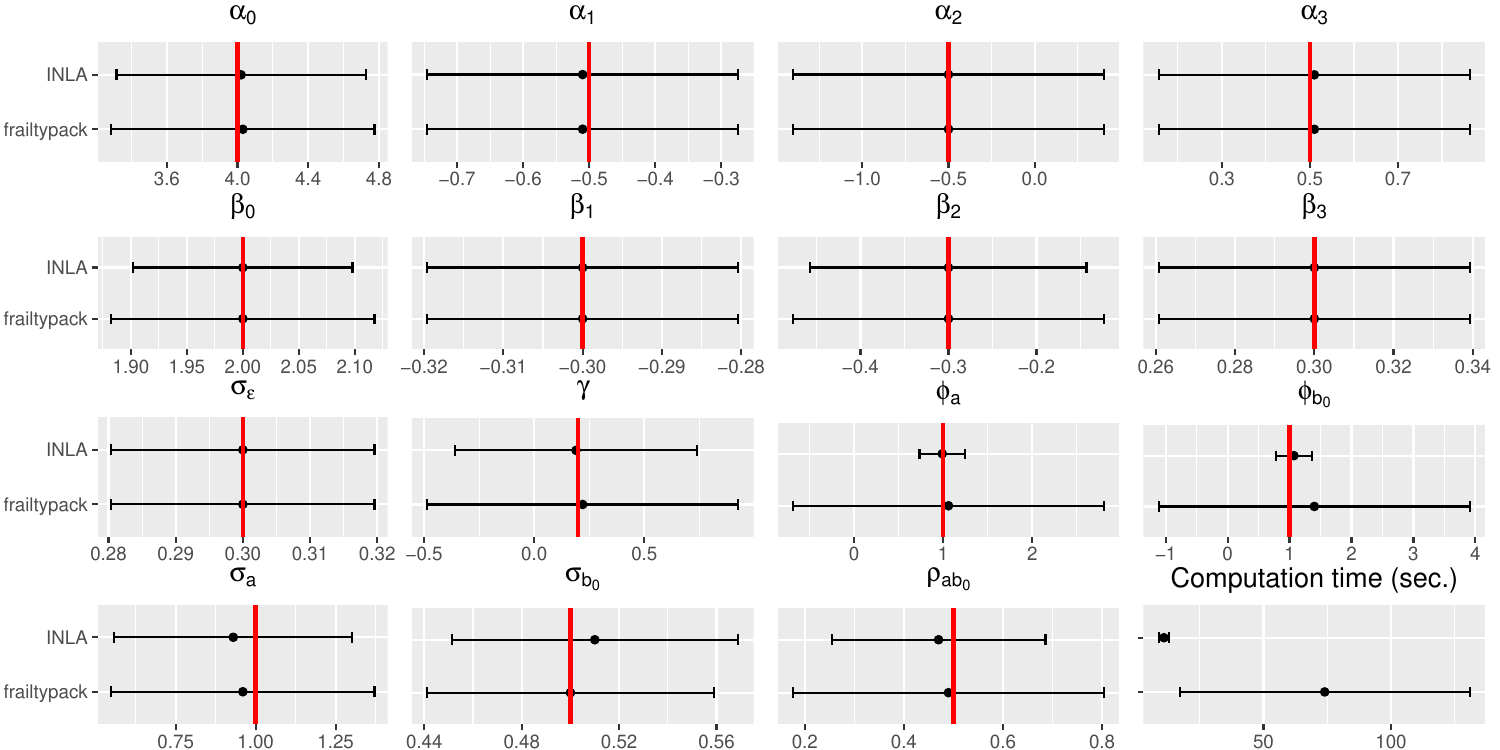}
\caption{Simulations with two correlated random effects (the black dot is the mean value with 95\% confidence intervals and true value is the red vertical bar)}
\label{Fig1}
\end{figure}

\subsubsection{Scenario 1: Two correlated random effects}
The results are displayed in Figure \ref{Fig1} and detailed in Table S1 of the supplementary materials. The fixed effect parameters from the binary and continuous parts are properly estimated with both algorithms, with similar precision and coverage probabilities close to the expected $95\%$ level. The parameter for the treatment effect in the survival part ($\gamma_1=0.2$) is associated to a larger variability with \textbf{frailtypack} ($\hat\gamma_1=0.22$, SD=$0.36$, CP=$96\%$) compared to \textbf{INLA} ($\hat\gamma_1=0.19$, SD=$0.28$, CP=$95\%$). The true value of the standard deviation of the random intercept in the binary part ($\sigma_a=1$) is within the $95\%$ credible interval with \textbf{INLA} ($\hat\sigma_a=0.93$, SD=$0.19$, CP=$92\%$), with a slightly lower posterior mean value compared to \textbf{frailtypack}'s estimate ($\hat\sigma_a=0.96$, SD=$0.21$, CP=$94\%$). The random intercept's standard deviation in the continuous part is found similar with both algorithms with a slightly lower coverage probability of the true value with \textbf{frailtypack} ($87\%$) but the correlation between the random intercepts of the binary and continuous parts ($\rho_{ab}=0.5$) has a reduced variability estimate with \textbf{INLA} ($\hat{\rho}_{ab}=0.47$, SD=$0.11$, CP=$94\%$) compared to \textbf{frailtypack} ($\hat{\rho}_{ab}=0.49$, SD=$0.16$, CP=$93\%$). The main difference observed is the estimation of the parameters for the association of the random effects with the risk of event, which links the biomarker to the terminal event. The association involving the random intercept from the binary part ($\varphi_a=1$) has much lower variability with \textbf{INLA} ($\hat\varphi_a=0.99$, SD=$0.13$, CP=$99\%$) and is unbiased with good coverage with \textbf{frailtypack} ($\hat\varphi_a=1.06$, SD=$0.89$, CP=$94\%$). The association involving the random intercept from the continuous part ($\varphi_b=1$) is biased upwards with \textbf{frailtypack} with large variability ($\hat\varphi_b=1.40$, SD=$1.28$, CP=$92\%$), while \textbf{INLA}'s posterior estimate recovers the true value ($\hat\varphi_b=1.07$, SD=$0.15$, CP=$98\%$). This could be due to the small sample size problems that cause more convergence issues under the frequentist framework. Although \textbf{INLA} yields accurate posterior estimates with small variability for these parameters, the coverage probabilities are higher than the expected $95\%$. The computation times are much lower with \textbf{INLA} ($6$ seconds per model, SD=$1$) compared to \textbf{frailtypack} ($74$ seconds per model, SD=$29$). Finally, $9\%$ of the $1000$ models did not reach convergence with \textbf{frailtypack}.

\begin{figure}[!ht]
\centering
\includegraphics[width=17cm, height=10cm]{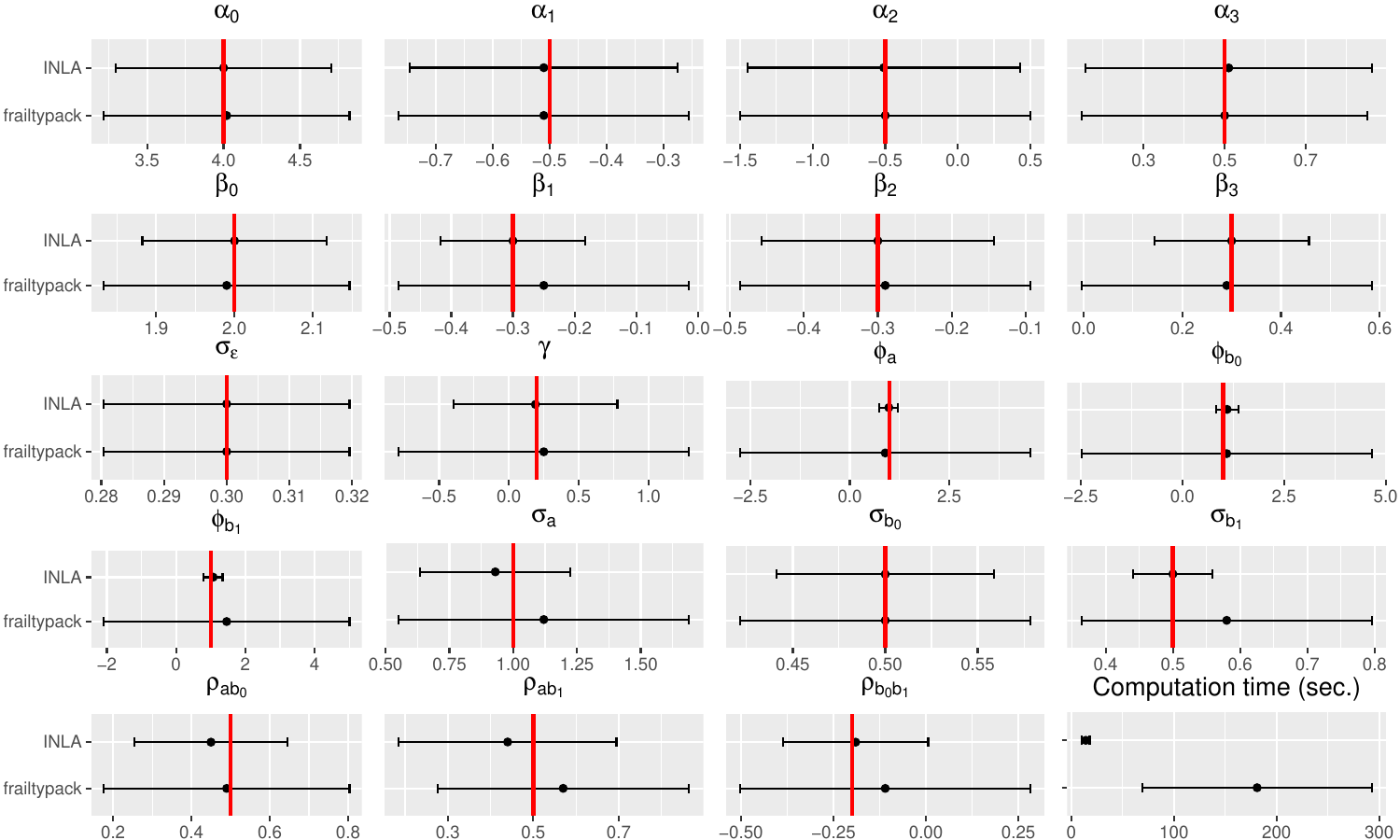}
\caption{Simulations with three correlated random effects (the black dot is the mean value with 95\% confidence intervals and true value is the red vertical bar)}
\label{Fig2}
\end{figure}

\subsubsection{Scenario 2: Three correlated random effects}
The results are displayed in Figure \ref{Fig2} and detailed in Table S2 of the supplementary materials. With an additional random effect parameter compared to scenario 1, the fixed effects parameters are still properly estimated with \textbf{INLA}. The coverage probabilities are low with \textbf{frailtypack} for the slope and treatment by slope parameters in the continuous part ($\beta_1=-0.3$ and $\beta_3=0.3$), while the parameter estimates remain unbiased. The variability for these two parameters is lower with \textbf{INLA} ($\hat\beta_1=-0.30$, SD=$0.06$, CP=$94\%$ and $\hat\beta_3=0.30$, SD=$0.08$, CP=$95\%$) compared to \textbf{frailtypack} ($\hat\beta_1=-0.25$, SD=$0.12$, CP=$44\%$ and $\hat\beta_3=0.29$, SD=$0.15$, CP=$47\%$). As observed in the first scenario, the treatment effect's posterior estimate in the survival model has lower variability with \textbf{INLA} ($\hat\gamma_1=0.19$, SD=$0.30$, CP=$94\%$), moreover the coverage probability for this parameter is lower than expected with \textbf{frailtypack} ($\hat\gamma_1=0.25$, SD=$0.53$, CP=$82\%$). For the random effects covariance structure estimation, the posterior mean from \textbf{INLA} is slightly lower than the true value of the random intercept's standard deviation in the binary part ($\hat\sigma_a=0.93$, SD=$0.15$, CP=$95\%$) while \textbf{frailtypack}'s value is slightly higher than the true value ($\hat\sigma_a=1.12$, SD=$0.29$, CP=$91\%$). Overall, \textbf{INLA} has lower variability for the standard deviation and correlation terms and \textbf{frailtypack} has poor coverage for these parameters (e.g., $22\%$ for $\sigma_{b_1}$ and $23\%$ for $\rho_{b_0b_1}$). The association parameters ($\varphi_a=1$, $\varphi_{b_0}=1$, $\varphi_{b_1}=1$) are recovered well and have much lower variability with \textbf{INLA} ($\hat\varphi_a=0.98$, SD=$0.12$, CP=$99\%$, $\hat\varphi_{b_0}=1.10$, SD=$0.14$, CP=$98\%$, $\hat\varphi_{b_1}=1.07$, SD=$0.14$, CP=$98\%$) compared to \textbf{frailtypack} ($\hat\varphi_a=0.89$, SD=$1.86$, CP=$89\%$, $\hat\varphi_{b_0}=1.09$, SD=$1.82$, CP=$89\%$, $\hat\varphi_{b_1}=1.46$, SD=$1.81$, CP=$89\%$), but still with conservative coverage probabilities. Computation times remain much lower with \textbf{INLA} ($6$ seconds per model, SD=$1$) compared to \textbf{frailtypack} ($181$ seconds per model, SD=$57$) for which the time increased substantially when adding the third random effect. Moreover, the convergence rate of the model is reduced with \textbf{frailtypack} for this scenario ($81\%$), because the model complexity increased.

\subsubsection{Scenario 3: Similar to scenario 2 with increased sample size of n=500}
The results are displayed in Figure \ref{Fig3} and detailed in Table S3 of the supplementary materials. With $n=500$ instead of $n=200$, the standard deviations of the mean parameters are reduced overall. However, the coverage probabilities for the slope and treatment by slope parameters in the continuous part ($\beta_1=-0.3$ and $\beta_3=0.3$) remain low with \textbf{frailtypack} ($\hat\beta_1=-0.30$, SD=$0.08$, CP=$45\%$ and $\hat\beta_3=0.29$, SD=$0.10$, CP=$46\%$), while the parameter estimates are unbiased. \textbf{INLA} has a lower variability and better coverage probabilities for these parameters ($\hat\beta_1=-0.30$, SD=$0.04$, CP=$92\%$ and $\hat\beta_3=0.30$, SD=$0.05$, CP=$95\%$) compared to \textbf{frailtypack}. For the random intercept's standard deviation in the binary part ($\sigma_a=1$), the posterior mean from \textbf{INLA} is slightly closer to the true value ($\hat\sigma_a=0.96$, SD=$0.11$) compared to scenario 2 while \textbf{frailtypack}'s estimate remains unbiased with a larger variability ($\hat\sigma_a=1.01$, SD=$0.14$). The other parameters in the random effects covariance structure are well recovered with lower variability for the standard deviation and correlation terms overall with \textbf{INLA}. The association parameters ($\varphi_a=1$, $\varphi_{b_0}=1$, $\varphi_{b_1}=1$) are recovered similarly as in scenario 2 with \textbf{INLA} ($\hat\varphi_a=0.91$, SD=$0.17$, CP=$96\%$, $\hat\varphi_{b_0}=1.12$, SD=$0.18$, CP=$94\%$, $\hat\varphi_{b_1}=1.09$, SD=$0.15$, CP=$97\%$), while their variability and bias is reduced with \textbf{frailtypack} ($\hat\varphi_a=0.99$, SD=$1.02$, CP=$92\%$, $\hat\varphi_{b_0}=1.06$, SD=$0.96$, CP=$89\%$, $\hat\varphi_{b_1}=1.08$, SD=$1.00$, CP=$89\%$), but remains much higher compared to \textbf{INLA}. Note that the high coverage for the association parameters with \textbf{INLA} in the first two scenarios was likely explained by the small sample size, leading to a higher importance of the non informative priors and resulting in high coverage. In this scenario with an increased sample size, the coverage is getting closer to the nominal $95\%$ level. We observe an increased difference in computation times between \textbf{INLA} ($11$ seconds per model, SD=$1$) and \textbf{frailtypack} ($340$ seconds per model, SD=$89$) compared to scenario 2. Finally, the convergence rate of the model has improved with \textbf{frailtypack} under this scenario ($96\%$).

\begin{figure}[!ht]
\centering
\includegraphics[width=17cm, height=10cm]{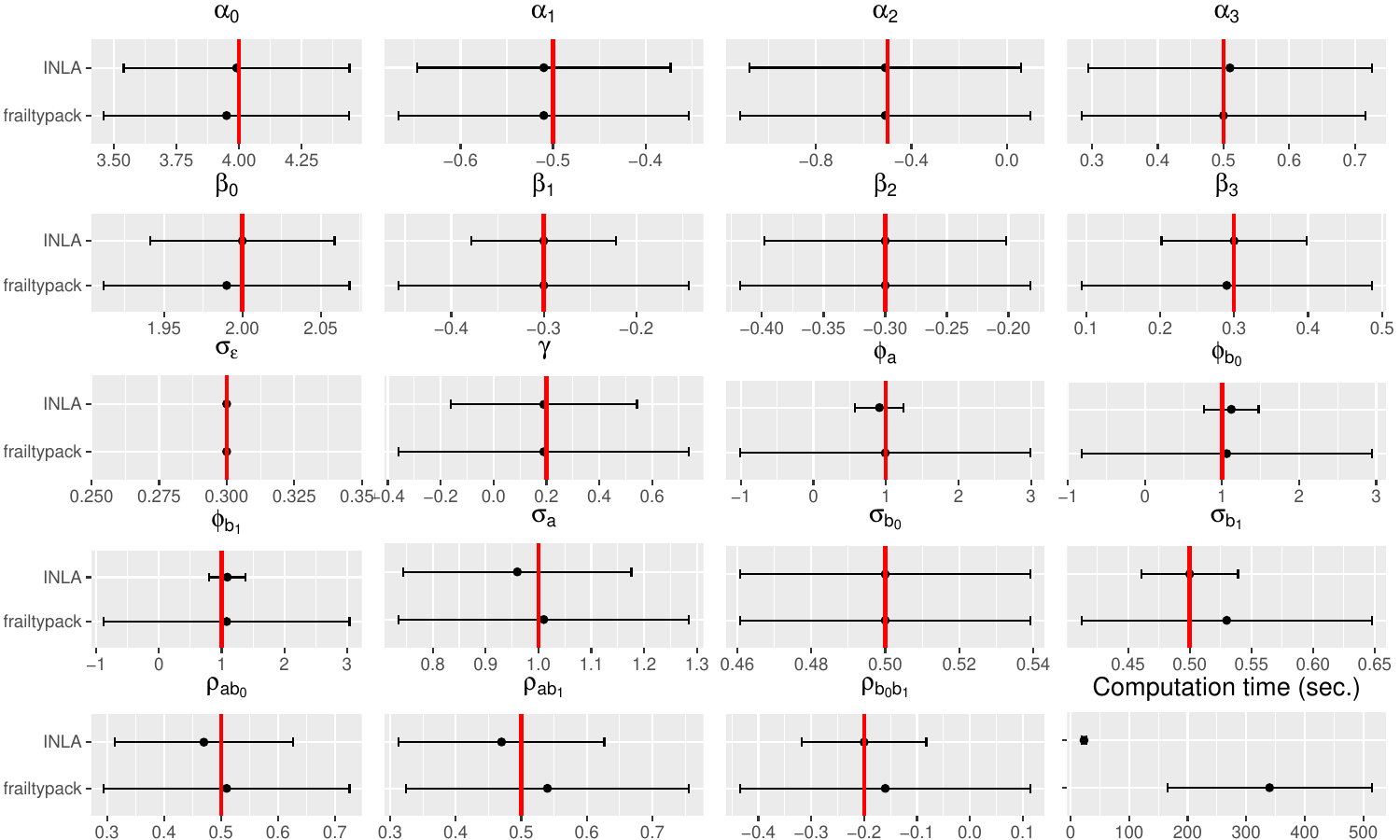}
\caption{Simulations with $n=500$ (the black dot is the mean value with 95\% confidence intervals and true value is the red vertical bar)}
\label{Fig3}
\end{figure}

\subsubsection{Scenario 4: Similar to scenario 2 with two natural cubic splines for the trend in the continuous part (4 correlated random effects, 200 individuals)}
The last scenario illustrates the behavior of \textbf{INLA} with a more flexible trend in the continuous part. It includes 4 correlated random effects and 200 individuals. The results are displayed in Figure \ref{Fig4} and detailed in Table S4 of the supplementary materials. The estimation with \textbf{frailtypack} had increased convergence issues compared to previous scenarios, $46\%$ of the 1000 models did not reach convergence. Moreover, fixed effects related to spline functions ($\beta_1=-1$, $\beta_2=-1$) have poor coverage and higher standard deviation with \textbf{frailtypack} ($\hat\beta_1=-1.00$, SD=$0.13$, CP=$88\%$, $\hat\beta_2=-1.02$, SD=$0.16$, CP=$85\%$) compared to \textbf{INLA} ($\hat\beta_1=-0.99$, SD=$0.09$, CP=$95\%$, $\hat\beta_2=-1.00$, SD=$0.09$, CP=$92\%$). Finally, the random effects variance and covariance parameters have higher variability and poorer coverage with \textbf{frailtypack} compared to \textbf{INLA}, in particular the random intercept from the binary part ($\sigma_a=0.5$) has a mean value off the true value ($\hat\sigma_a=0.78$, SD=$0.25$, CP=$89\%$) with \textbf{frailtypack} while \textbf{INLA} is recovering this parameter with more accuracy ($\hat\sigma_a=0.55$, SD=$0.10$, CP=$98\%$). Note that this scenario is associated to lower zero rate on average ($5\%$), which may explain the poor result of \textbf{frailtypack} for this parameter.

\begin{figure}[!ht]
\centering
\includegraphics[width=17cm, height=14cm]{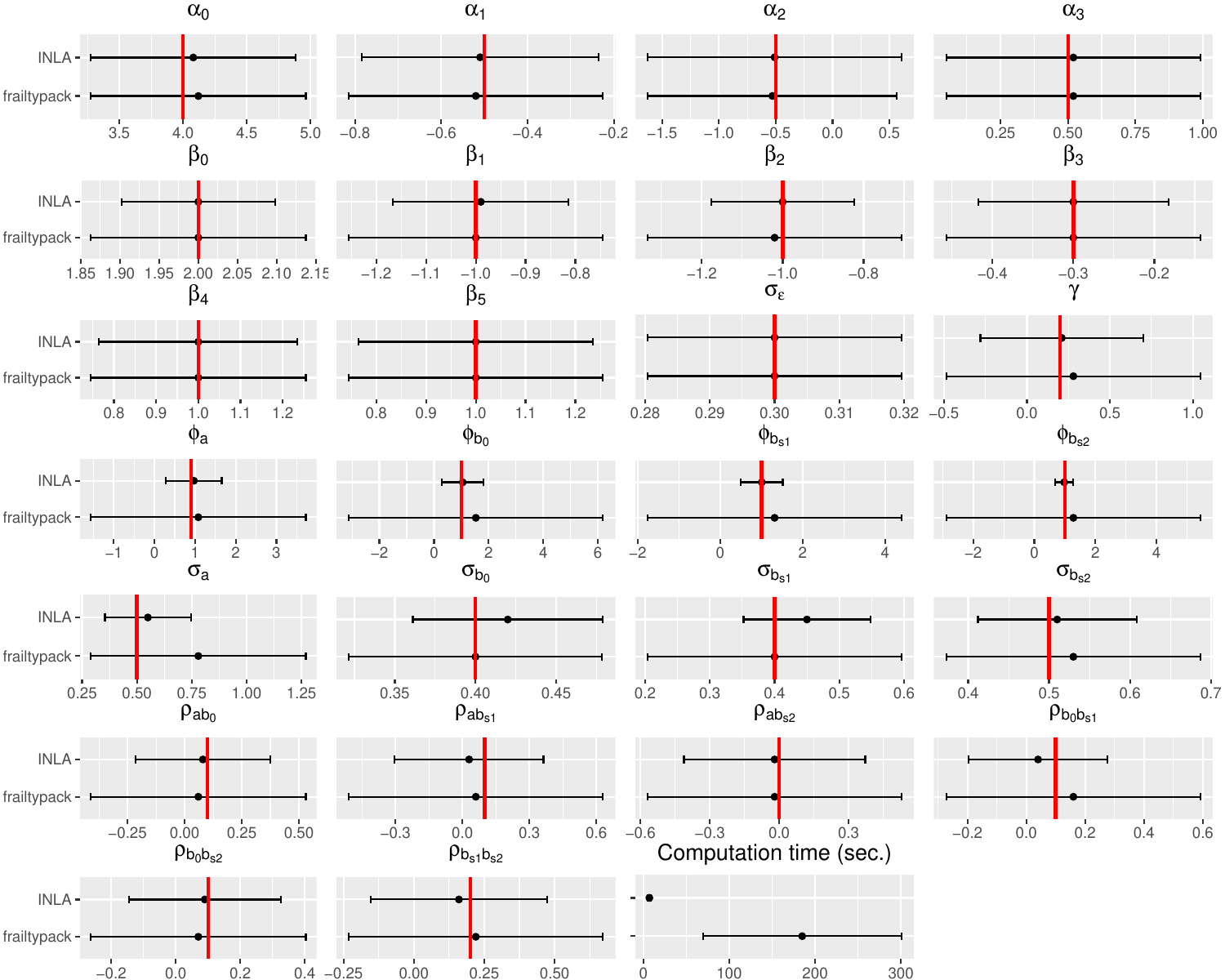}
\caption{Simulations with splines (the black dot is the mean value with 95\% confidence intervals and true value is the red vertical bar)}
\label{Fig4}
\end{figure}

\begin{figure}[!ht]
\centering
\includegraphics[width=17cm, height=5cm]{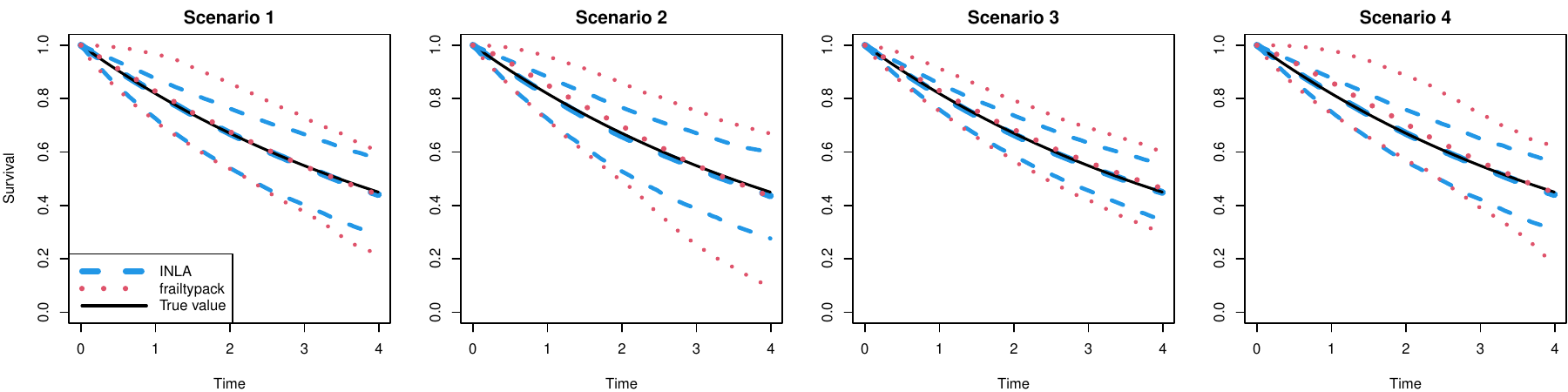}
\caption{Quantiles at $2.5\%$, $50\%$ and $97.5\%$ of the baseline survival curves estimated with \textbf{frailtypack} and \textbf{INLA} in the simulations (Section \ref{sec:sim}).}
\label{Fig5}
\end{figure}

\subsection{Conclusions}
Our method comparison suggests that the frequentist approach, implemented in \textbf{frailtypack}, reaches some limitations when fitting the more complex TPJMs, compared to the Bayesian approach implemented in \textbf{INLA}, which provides accurate approximations of posterior quantities. Convergence might be an issue and estimation of the association parameters is highly variable with \textbf{frailtypack}. A representation of the baseline survival curves estimated under both scenarios is displayed in Figure \ref{Fig5}. The median of the estimated survival curves is unbiased with both estimation strategies but \textbf{INLA} has a reduced variability compared to \textbf{frailtypack}. Note that \textbf{INLA} being a deterministic algorithm, the notion of ``convergence rate" does not apply and is only provided in the results for comparison with the iterative algorithm implemented in \textbf{frailtypack}. R codes including data simulation and model estimation with \textbf{frailtypack} and \textbf{INLA} assuming a conditional two-part joint model are available at $github.com/DenisRustand/TPJM\_sim$ and in the supplementary materials for scenario 2, other scenarios follow trivially.

\section{Application}
\label{sec:appli}
We applied the Bayesian TPJM to two cancer clinical trials, the GERCOR and the PRIME studies. A comparison with \textbf{frailtypack} is provided only for the GERCOR data since this approach did not converge on the PRIME study. We used the same parameterization for \textbf{INLA} and \textbf{frailtypack}, as detailed in the simulation studies but with an increased number of integration points for the Monte Carlo method in \textbf{frailtypack} (i.e., 5000 points). In the context of a Bayesian approximation of the MLE, we provide indications of the p-value for both \textbf{frailtypack} and \textbf{INLA} to ease the interpretation and the comparison of the results. These p-values are computed using the Z-score, assuming normal distributions. Each package has a specific criterion for model selection that takes into account the goodness-of-fit and the complexity of the model (i.e., number of parameters). \textbf{Frailtypack} uses the Likelihood-based Cross Validation (\citet{Commenges07}), which accounts for its penalized likelihood while \textbf{INLA} is based on the Deviance Information Criterion (\citet{Spiegelhalter02}).

\begin{table}[ht]
\caption{Description of the GERCOR and PRIME study datasets}
\centering
\scriptsize
{\tabcolsep=2.25pt
\begin{tabular}{@{}lllll@{}}
\hline
Study & \multicolumn{2}{c}{GERCOR} & \multicolumn{2}{c}{PRIME}\\
\hline
\multirow{2}{*}{Treatment} & arm A & arm B & arm A & arm B\\
& FOLFIRI/FOLFOX6 & FOLFOX6/FOLFIRI & FOLFOX4 & Panitumumab/FOLFOX4 \\
Number of patients enrolled & 109 & 111 & 593 & 590\\
Number of patients for the analysis & 101 & 104 & 223 & 219\\
number of repeated measurements of the SLD & 748 & 727 & 1192 & 1081 \\
Number of zero values (\%) & 118 (16.2\%) & 56 (7.5\%) & 47 (3.8\%) & 52 (4.6\%) \\
Number of death (\%) & 83 (82.2\%) & 82 (78.8\%) & 164 (73.5\%) & 164 (74.9\%)\\
Median OS (years) & 1.8 (1.4-2.3) & 1.8 (1.5-2.2) & 1.7 (1.5-1.9) & 1.4 (1.3-1.7)\\
\hline
KRAS exon 2 at codons 12 and 13 & & & &\\
\ Nonmutated & & & 132 (59.2\%) & 128 (58.4\%)\\
\ Mutated & & & 91 (40.8\%) & 91 (41.6\%)\\
\ Not available & 101 (100\%) & 104 (100\%) & &\\
\end{tabular}}
\label{descAppli}
\end{table}

\subsection{GERCOR study}
\subsubsection{Description}
It is a randomized clinical trial investigating two treatment strategies that included a total of 220 patients with metastatic colorectal cancer. The reference strategy (arm A) corresponds to FOLFIRI (irinotecan) followed by FOLFOX6 (oxaliplatin) while arm B involves the reverse sequence. Patients were randomly assigned from December 1997 to September 1999 and the date chosen to assess overall survival was August 30, 2002. Complete data are available on 205 individuals for data analysis. Among them, 165 (80\%) died during the follow-up. There are 1475 repeated measurements for the biomarker, 174 of which are zero values (12\%). A summary of the dataset structure is given in Table \ref{descAppli}. Our model uses death as the terminal event and the repeated SLD measurements (in centimeters) as the semicontinuous biomarker. Additional baseline covariates collected at the start of the study are also included, including performance status (0/1/2), lung metastatic site (Y/N), previous adjuvant radiotherapy (Y/N), previous surgery (no surgery/curative/palliative) and metastases (metachronous/synchronous). The first analysis of this dataset (\citet{Tournigand04}) did not find any significant difference between the two treatment strategies using classic survival analysis methods (i.e. log-rank tests). A trivariate joint model has been applied to this study for the simultaneous analysis of the longitudinal SLD, recurrent events (progression of lesions not included in the SLD or new lesions) and the terminal event (\citet{Krol18}). A flexible mechanistic model using ordinary differential equation was proposed to fit the biomarker dynamics. The results show a greater decline of the SLD for treatment arm A compared to treatment arm B. Moreover, the model finds a strong association between the biomarker model and the risk of terminal event. However the interpretation of this treatment effect is difficult due to the non-linear transformation applied to the outcome (Box-Cox) and the use of a non-linear mechanistic model. Finally, a conditional two-part joint model was recently proposed (\citet{Rustand20}), which showed a significant treatment effect on the positive values of the biomarker (and no treatment effect on the probability of zero value). The model was able to show that when taking into account this treatment effect on the biomarker, the risk of terminal event is not significantly different between the two treatment arms. In the results, the mean parameters and their standard deviation are obtained by taking the ML estimates and the inverse Hessian matrix with \textbf{frailtypack} while the posterior mean and standard deviation of the posterior distribution were used with \textbf{INLA}. For the evolution of the SLD over time in the continuous part, we compared a model with a linear trend (fixed + random slope) with a more flexible model with two natural cubic splines with a knot at the median of observed times which corresponds to 6 months of follow-up (each spline is associated to a fixed and a random effect). The goodness of fit criterion was in favor of the flexible model with both \textbf{INLA} (DIC = 2787 with linear trend and DIC = 2260 with splines) and \textbf{frailtypack} (LCV = 0.45 with linear trend and LCV = 0.40 with splines). The splines are described in Figure S1 of the supplementary materials and the results with a linear trend are given in Table S5 of the supplementary materials. The first spline corresponds to an initial increase during the first year followed by a decrease for the rest of the follow-up while the second spline has an initial decrease during the first year and then increases for the rest of the follow-up.

\begin{table}[ht]
\caption{Application of the Bayesian and frequentist two-part joint models with two natural cubic splines to the GERCOR study with \textbf{INLA} and \textbf{frailtypack}}
\centering
\scriptsize
{\tabcolsep=2.25pt
\begin{tabular}{@{}llllll@{}}
\hline
Approach & & INLA & frailtypack\\
& & Est.$^\dagger$ (SD$^\ddagger$) & Est. (SD)\\
\hline
\textbf{Binary part} (SLD$>$0 versus SLD=0) & & & \\
\ intercept & $\alpha_0$ & 5.37*** (0.62) & 5.72*** (0.71) \\
\ time (year) & $\alpha_1$ & -2.01*** (0.37) & -2.02*** (0.40)\\
\ treatment (B/A) & $\alpha_2$ & -0.98 (0.72) & -1.35 (0.69) \\
\ PS (1 vs. 0) & $\alpha_3$ & 2.03*** (0.57) & 2.17*** (0.57)\\
\ PS (2 vs. 0) & $\alpha_4$ & 1.56 (1.12) & 1.70 (1.14) \\
\ previous\_radio (Y/N) & $\alpha_5$ & 0.71 (0.71) & 0.59 (0.71) \\
\ lung (Y/N) & $\alpha_6$ & 1.81** (0.65) & 1.54* (0.62) \\
\ time:treatment (B/A) & $\alpha_7$ & 0.31 (0.44) & 0.36 (0.47) \\
\textbf{Continuous part} ($\textrm{E}[\log(Y_{ij})|Y_{ij}>0]$) & & & \\
\ intercept & $\beta_0$ & 2.26*** (0.13) & 2.15*** (0.09) \\
\ slope 1 & $\beta_1$ & -0.26 (0.20) & 0.06 (0.15) \\
\ slope 2 & $\beta_2$ & 1.48*** (0.37) & 1.93*** (0.29) \\
\ treatment (B/A) & $\beta_3$ & -0.23** (0.09) & -0.23** (0.07) \\
\ PS (1 vs. 0) & $\beta_4$ & 0.35*** (0.09) & 0.41*** (0.07) \\
\ PS (2 vs. 0) & $\beta_5$ & 0.40** (0.14) & 0.46*** (0.10) \\
\ previous\_surgery (curative) & $\beta_6$ & -0.47** (0.16) & -0.26* (0.11) \\
\ previous\_surgery (palliative) & $\beta_7$ & -0.03 (0.12) & 0.10 (0.09)\\
\ previous\_radio (Y/N) & $\beta_8$ & -0.22 (0.10) & -0.23 (0.08) \\
\ metastases (metachronous vs. synchronous) & $\beta_9$ & 0.35* (0.14) & 0.25* (0.10) \\
\ slope 1:treatment (B/A) & $\beta_{10}$ & 0.91** (0.28) & 0.53* (0.24) \\
\ slope 2:treatment (B/A) & $\beta_{11}$ & 1.49*** (0.54) & 0.85* (0.39) \\
\ residual S.E.  & $\sigma_\varepsilon$  & 0.25*** (0.01) & 0.25*** (0.01) \\
\hline
\textbf{Death risk} & & & \\
\ treatment (B/A) & $\gamma_1$ & 0.25 (0.24) & 0.07 (0.23)\\
\ PS (1 vs. 0) & $\gamma_2$ & 0.94*** (0.22) & 1.05*** (0.24) \\
\ PS (2 vs. 0) & $\gamma_3$ & 1.66*** (0.36) & 1.79*** (0.39) \\
\ previous\_surgery (curative) & $\gamma_4$& -0.95* (0.44) & -0.70 (0.45) \\
\ previous\_surgery (palliative) & $\gamma_5$ & -0.50 (0.31) & -0.39 (0.32) \\
\ metastases (metachronous vs. synchronous) & $\gamma_6$& 0.83* (0.36) & 0.71 (0.37) \\
\textbf{Association} & & & \\
\ intercept (binary part) & $\varphi_a$ & 0.05 (0.08) & -0.07 (0.15) \\
\ intercept (continuous part) & $\varphi_{b_0}$ & 0.98*** (0.22) & 1.26* (0.53) \\
\ slope 1 (continuous part) & $\varphi_{b_{s1}}$& 0.57*** (0.17) & 0.76** (0.29) \\
\ slope 2 (continuous part) & $\varphi_{b_{s2}}$& 0.06 (0.07) & 0.02 (0.08) \\
\hline
\textbf{Random effects's standard deviation} & & \\
\ intercept (binary part) & $\sigma_{a}$ & 3.08*** (0.37) & 3.12*** (0.39)\\
\ intercept (continuous part) & $\sigma_{b_0}$ & 0.59*** (0.04) & 0.57*** (0.03)\\
\ slope 1 (continuous part) & $\sigma_{b_{s1}}$ & 1.52*** (0.15) & 1.73*** (0.12)\\
\ slope 2 (continuous part) & $\sigma_{b_{s2}}$ & 2.50*** (0.29) & 2.97*** (0.25)\\
\ & $\rho_{a b_0}$ & 0.51*** (0.08) & 0.53*** (0.06)\\
\ & $\rho_{a b_{s1}}$ & 0.59*** (0.12) & 0.70*** (0.07)\\
\ & $\rho_{a b_{s2}}$ & 0.32* (0.16) & 0.55*** (0.09)\\
\ & $\rho_{b_0 b_{s1}}$ & -0.14 (0.10) & -0.03 (0.05)\\
\ & $\rho_{b_0 b_{s2}}$ & -0.09 (0.12) & 0.08 (0.05)\\
\ & $\rho_{b_{s1} b_{s2}}$ & 0.72 (0.08)*** & 0.78*** (0.03)\\
\hline
\multicolumn{4}{l}{\textbf{Computation time (Intel Xeon Gold 6248 2.50GHz)}} \\
\ 8 CPUs & & 13 sec. & 11230 sec.\\
\ 80 CPUs & & 10 sec. & 2355 sec.\\
\hline
\vspace{-0.2cm}\\
\multicolumn{4}{l}{$^\dagger$ Posterior mean, $^\ddagger$ Posterior standard deviation}, \textsuperscript{***}$p<0.001$, 
  \textsuperscript{**}$p<0.01$, 
  \textsuperscript{*}$p<0.05$\\
\end{tabular}}
\label{resGERCOR}
\end{table}

\begin{figure}[ht]
\centering
\includegraphics[scale=0.8]{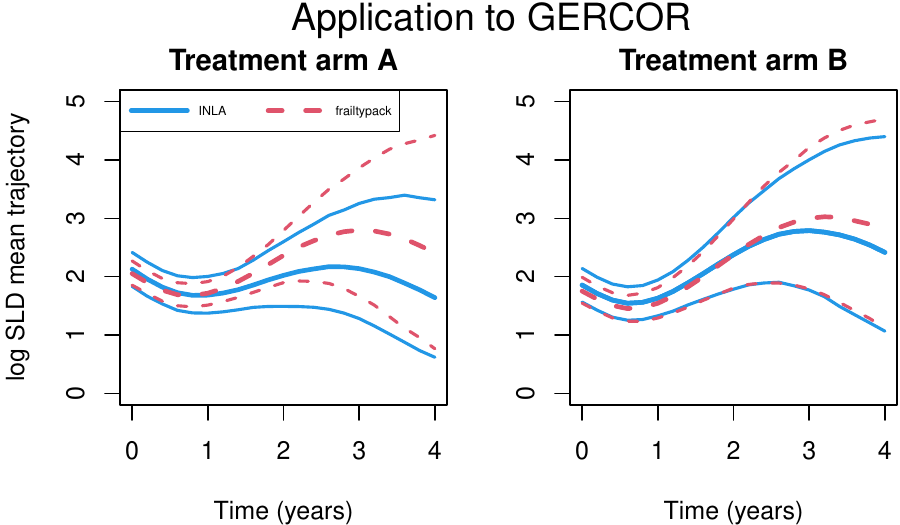}
\caption{Mean biomarker value according to treatment received. The 95\% credible intervals are obtained by resampling from the posterior parameter distributions.}
\label{Fig6}
\end{figure}

\begin{figure}[ht]
\centering
\includegraphics[scale=0.8]{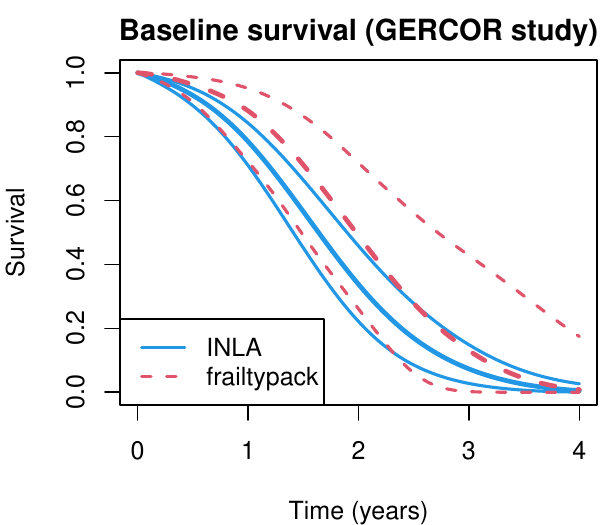}
\caption{Baseline survival curves and their 95\% confidence and credible intervals obtained from the application of the TPJM to the GERCOR study with \textbf{frailtypack} and \textbf{INLA}, respectively.}
\label{Fig7}
\end{figure}

\subsubsection{Results}
As presented in Table \ref{resGERCOR}, the fixed effect parameter estimates in the binary, continuous and survival parts are quite similar between the frequentist and Bayesian approaches. The evolution of the SLD over time conditional on treatment is displayed in Figure \ref{Fig6}, where random effects have been integrated out to have population average trajectories. Both treatment arms have a similar evolution of the log SLD over time. The uncertainty is slightly reduced with \textbf{INLA} at the end of the follow-up compared to \textbf{frailtypack}. Moreover, the mean value of the log SLD with \textbf{INLA} gets slightly lower over time compared to \textbf{frailtypack}, but no significant difference is observed marginally despite a significant association between the second spline and treatment with \textbf{INLA} ($\hat\beta_{11}=1.49$, SD=$0.54$), while \textbf{frailtypack} finds a lower effect size but still significant ($\hat\beta_{11} = 0.85$, SD=$0.39$). This effect of treatment indicates that treatment arm B may be associated to a stronger reduction of the log SLD in early follow-up followed by a higher increase of the log SLD in the late follow-up compared to treatment arm A. This difference among positive values does not translate into a significant difference in the marginal evolution of the log SLD (i.e., including both zeros and positives), suggesting that models mixing zeros and positive (e.g., Tobit model) may miss this effect. Figure S2 displays observed vs. fitted longitudinal trajectory of the log SLD for five patients representative of the dataset, illustrating the good fit of the model. These fitted trajectories are only displayed with \textbf{INLA} since the linear predictors for each observations are directly available as part of the model's output while \textbf{frailtypack} does not provide such information.

The hazard ratio of treatment arm B versus treatment arm A that evaluates the change in the risk of death was higher with \textbf{INLA} (HR=$1.28$, CI $0.80-2.06$) compared to \textbf{frailtypack} (HR=$1.07$, CI $0.68-1.69$), but not significant in both cases. The main difference between \textbf{INLA} and \textbf{frailtypack} is in the estimation of the parameters for the association between the two-part model for the biomarker and the survival model. There is a positive and significant association between the random intercept ($\hat\varphi_{b_0}=0.98$, SD=$0.22$) and the first spline ($\hat\varphi_{b_{s1}}=0.57$, SD=$0.17$) from the continuous part and the risk of event with \textbf{INLA}. This association has a higher effect size and much larger variability with \textbf{frailtypack} ($\hat\varphi_{b_0}=1.26$, SD=$0.52$ and $\hat\varphi_{b_1}=0.76$, SD=$0.29$). This is in line with our simulation results (scenario 1) where the association structure was estimated with better precision with \textbf{INLA}. The computation time for \textbf{frailtypack} increases quickly with the sample size and the model complexity (number of parameters and dimension of the random effects). The model was estimated in 11230 seconds with \textbf{frailtypack} with 8 CPUs and this reduces to 2355 seconds with 80 CPUs while it is estimated in less than 15 seconds with \textbf{INLA}. The differences found in the association structure estimates is important when assessing the relationship between the biomarker dynamics and the risk of event. For instance, let's assume a clinician is interested in the top 15\% patients who had the largest SLD at baseline compared to the average patient. Their random effect $b_{0i}$ should be higher than 1 standard deviation, that is from Table \ref{resGERCOR}, $b_{0i} > 0.59$ with \textbf{INLA} (respectively $b_{0i} > 0.57$ with \textbf{frailtypack}). Conditional on $b_{0i} > 0.59$ (respectively $b_{0i} > 0.57$), the mean values of the random effects can be derived by sampling from a conditional multivariate normal distribution with correlation matrix given in Table \ref{resGERCOR}. These conditional means are $2.38$, $0.90$, $-0.28$ and $-0.25$ for $a$, $b_0$, $b_{s1}$ and $b_{s2}$, respectively ($2.51$, $0.87$, $-0.07$ and $0.35$ with \textbf{frailtypack}). Therefore, these top 15\% individuals increase their chance to have the terminal event (i.e., to die) measured by an hazard ratio of HR = $\exp(0.05*2.38+0.98*0.90+0.57*(-0.28)+0.06*(-0.25)) = \exp(0.83) = 2.29$, CI=$2.15-2.40$, compared to a patient who has an average longitudinal SLD profile. \textbf{Frailtypack} gives a similar hazard ratio but with higher uncertainty HR = $\exp((-0.07)*2.51+1.26*0.87+0.76*(-0.07)+0.02*0.35) = \exp(0.87) = 2.40$, CI=$1.68-3.30$. The confidence intervals were obtained by sampling parameters from the inverse Hessian matrix with \textbf{frailtypack} and the posterior distribution of the parameters with \textbf{INLA}. Figure \ref{Fig7} shows the baseline survival curves obtained with \textbf{frailtypack} and \textbf{INLA}, the survival decreases faster with \textbf{INLA} and has a reduced uncertainty compared to \textbf{frailtypack}, although no significant differences is observed between the two estimation strategies.

\subsection{PRIME study}
\subsubsection{Description}
The Panitumumab Randomized Trial in Combination with Chemotherapy for Metastatic Colorectal Cancer to Determine Efficacy (PRIME) study is a more challenging application for fitting the TPJM because it includes information about the KRAS mutation status (exon 2 codons 12/13), which has been shown to impact the clinical response to treatment in metastatic colorectal cancer patients (\citet{VanCutsem08, Normanno09, Bokemeyer08}). It is therefore an important risk modifier and clinicians are interested to assess treatment by mutation interaction in order to tailor treatment to patients' genetic risk (\citet{Marabelle20}). This dataset is freely available on ProjectDataSphere.org (PDS UID: Colorec\_Amgen\_2006\_309).

The PRIME study is a randomized clinical trial that compares the efficacy and safety of panitumumab (anti-EGFR) in combination with FOLFOX4 (chemotherapy) with those of FOLFOX4 alone in the first-line treatment of patients, according to KRAS exon 2 status (Wild type or Mutant type). Between August 2006 and February 2008, 1183 patients were randomly assigned to receive treatment arm A (FOLFOX4 alone) or treatment arm B (panitumumab + FOLFOX4). The data for analysis includes a subset of 442 patients (i.e., 741 excluded from the publicly available dataset). There are 2372 repeated measurements of the SLD, 99 of which are zero values (4\%). The small rate of zero measurements in the SLD distribution leads to a large variability in the binary part, however zeros corresponds to patients with a complete shrinkage of their target lesions, which is a very relevant information for clinicians about treatment effect. The number of individual repeated measurements for this biomarker varies between 1 and 24 with a median of 5. The death rate is 74\%, corresponding to 328 deaths. Summary statistics of the dataset are given in Table \ref{descAppli}. Additional baseline covariates collected at the start of the study are also included, including metastases to liver at study entry (Y/N), the number of baseline metastases sites (1/2/3/4+), age ($<$60/60-69/$>=$70) and baseline ECOG performance status (0/1/2). We used a global backward selection procedure for each component of the model to select the covariates to include in the final joint model. The conclusions of the study are presented in \citet{Douillard13} and show the importance of taking into account the mutation status when assessing treatment effect. Among patients without mutated KRAS, treatment arm B was associated with a slightly significant reduced risk of death compared to treatment arm A. For patients with mutated KRAS, treatment arm B was associated with a non-significant increase in the risk of death compared to treatment arm A. Unlike the first application, we assume a linear trend on the log scale for the continuous part of the TPJM because the model with spline functions was not fitting well the data (high variability of the posterior parameter estimates that reflects the non informative priors), suggesting that the data is not informative enough to fit all the parameters properly (small sample size, short follow-up due to high death rate and many added parameters due to interactions between splines, treatment and KRAS mutation status).

\begin{table}[!ht]
\caption{Application of the Bayesian two-part joint model with shared random effects to the PRIME study with the R package \textbf{R-INLA}}
\centering
\scriptsize
{\tabcolsep=2.25pt
\begin{tabular}{@{}lllll@{}}
\hline
Approach & & R-INLA \\
& & Est.$^\dagger$ (SD$^\ddagger$) \\
\hline
\textbf{Binary part} (SLD$>$0 versus SLD=0) & &\\
\ intercept & $\alpha_0$ & 16.50*** (3.49) \\
\ time (year) & $\alpha_1$ & -8.31*** (1.66) \\
\ treatment (B/A) & $\alpha_2$ & 3.15 (3.68) \\
\ kras (MT/WT) & $\alpha_3$ & 6.18 (4.54) \\
\ treatment (B/A):kras (MT/WT) & $\alpha_4$ & -1.26 (6.73) \\
\ time:treatment (B/A) & $\alpha_5$ & 1.27 (1.81)\\
\ time:kras (MT/WT) & $\alpha_6$ & 0.64 (2.36) \\
\ time:treatment (B/A):kras (MT/WT) & $\alpha_7$ & 0.73 (4.42) \\
\textbf{Continuous part} $\textrm{E}[\log(Y_{ij})|Y_{ij}>0]$ & &\\
\ intercept & $\beta_0$ & 2.55*** (0.17)\\
\ time (years) & $\beta_1$ & -1.64*** (0.10) \\
\ treatment (B/A) & $\beta_2$ & -0.22* (0.10) \\
\ kras (MT/WT) & $\beta_3$ & -0.24* (0.11) \\
\ liver metastases (Y/N) & $\beta_4$ & 0.56*** (0.13) \\
\ ECOG (symptoms but ambulatory vs. fully active) & $\beta_5$ & 0.18* (0.07) \\
\ ECOG (in bed less than 50\% of the time vs. fully active) & $\beta_6$ & 0.49** (0.17) \\
\ baseline metastases sites (2 vs. 1) & $\beta_7$ & 0.08 (0.11) \\
\ baseline metastases sites (3 vs. 1) & $\beta_8$ & 0.23* (0.11) \\
\ baseline metastases sites (4+ vs. 1)  & $\beta_9$ & 0.18 (0.12) \\
\ treatment (B/A):kras (MT/WT) & $\beta_{10}$ & 0.23 (0.15) \\
\ time:treatment (B/A) & $\beta_{11}$ & 0.93*** (0.14) \\
\ time:kras (MT/WT) & $\beta_{12}$ & 1.18*** (0.15) \\
\ time:treatment (B/A):kras (MT/WT) & $\beta_{13}$ & -1.12*** (0.20) \\
\ residual S.E. & $\sigma_\varepsilon$ & 0.27*** (0.01) \\
\hline
\textbf{Death risk} & & \\
\ treatment (B/A) & $\gamma_1$ & 0.08 (0.17)\\
\ kras (MT/WT) & $\gamma_2$ & 0.18 (0.18) \\
\ treatment (B/A):kras (MT/WT) & $\gamma_3$ & 0.07 (0.24) \\
\ age (60-69 vs. $<$60) & $\gamma_4$ & 0.10 (0.13) \\
\ age (70+ vs. $<$60) & $\gamma_5$& 0.25 (0.14) \\
\ liver metastases (Y/N) & $\gamma_6$ & 0.05 (0.23) \\
\ ECOG (symptoms but ambulatory vs. fully active) & $\gamma_7$ & 0.31* (0.12) \\
\ ECOG (in bed less than 50\% of the time vs. fully active) & $\gamma_8$& 0.85** (0.26) \\
\ baseline metastases sites (2 vs. 1) & $\gamma_9$& 0.13 (0.20) \\
\ baseline metastases sites (3 vs. 1) & $\gamma_{10}$& 0.33 (0.20) \\
\ baseline metastases sites (4+ vs. 1) & $\gamma_{11}$& 0.44* (0.21) \\
\textbf{Association} & &\\
\ intercept (binary part) & $\varphi_a$ & 0.00 (0.01)\\
\ intercept (continuous part) & $\varphi_{b_0}$ & 0.47*** (0.10) \\
\ slope (continuous part) & $\varphi_{b_1}$ & 0.10 (0.14) \\
\hline
\textbf{Random effects's standard deviation} & &\\
\ intercept (binary part) & $\sigma_{a}$ & 8.61\\
\ intercept (continuous part) & $\sigma_{b_0}$ & 0.73\\
\ slope (continuous part) & $\sigma_{b_1}$ & 0.73\\
\ & $\rho_{a b_0}$ & 0.00\\
\ & $\rho_{a b_1}$ & 0.76\\
\ & $\rho_{b_0 b_1}$ & -0.23\\
\hline
\multicolumn{3}{l}{\textbf{Computation time (Intel Xeon E5-4627 v4 2.60 GHz)}} \\
\ 8 CPUs & & 46 sec.\\
\ 80 CPUs & & 39 sec.\\
\hline
\vspace{-0.2cm}\\
\multicolumn{3}{l}{$^\dagger$ Posterior mean, $^\ddagger$ Posterior standard deviation}, \textsuperscript{***}$p<0.001$, 
  \textsuperscript{**}$p<0.01$, 
  \textsuperscript{*}$p<0.05$\\
\end{tabular}}
\label{resPRIME}
\end{table}

\subsubsection{Results}
As presented in Table \ref{resPRIME}, in the binary part of the TPJM, the intercept is very large ($\hat\alpha_0=16.50$, SD=$3.49$), corresponding to a high probability of positive value at baseline. This probability is increased for patients with mutated KRAS ($\hat\alpha_3=6.18$, SD=$4.54$) and patients receiving treatment arm B ($\hat\alpha_2=3.15$, SD=$3.68$) but with large standard deviations so that these effects are not significant. The slope parameter with time is negative and significant ($\hat\alpha_1=-8.31$, SD=$1.66$), meaning that patients without mutated KRAS and receiving treatment A have a higher odds of zero SLD value over time, i.e., complete response to treatment. This odds decreases, but not significantly, among patients with either mutated KRAS ($\hat\alpha_6=0.64$, SD=$2.36$) or receiving treatment B ($\hat\alpha_5=1.27$, SD=$1.81$) and in patients with both mutated KRAS and receiving treatment arm B ($\hat\alpha_7=0.73$, SD=$4.42$).

In the continuous part of the TPJM, patients with the wild type KRAS status and in treatment arm A are associated with a decrease in the SLD value over time conditional on a positive SLD value ($\hat\beta_1=-1.64$, SD=$0.10$). This reduction of SLD over time is attenuated in patients with mutated KRAS ($\hat\beta_{12}=1.18$, SD=$0.15$) or receiving treatment B ($\hat\beta_{11}=0.93$, SD=$0.14$). Patients with the KRAS mutation and who received treatment B have a similar SLD trend over time as patients with the KRAS mutation who received treatment A or patients who received treatment B but with the wild type KRAS status because of the negative interaction term between time, treatment and KRAS status ($\hat\beta_{13}=-1.12$, SD=$0.20$).

In the survival part, the model shows no significant difference between treatment arms for the risk of death ($\hat\gamma_1=0.08$, SD=$0.17$). Besides, patients with mutated KRAS have similar risk of death compared to patients with the wild type ($\hat\gamma_2=0.18$, SD=$0.18$), so do patients with mutated KRAS receiving treatment B ($\hat\gamma_3=0.07$, SD=$0.24$). The random effect from the binary part and the random slope from the continuous part are not associated to the risk of death ($\hat\varphi_a=0.00$, SD=$0.01$ and $\hat\varphi_{b1}=0.10$, SD=$0.14$) but the random intercept from the continuous part ($\hat\varphi_{b0}=0.47$, SD=$0.10$) has a positive and highly significant association with the risk of event. This means that conditional on a positive value, the individual deviation from the mean baseline value of the SLD is predictive of the risk of event. Similarly to the GERCOR study, we can compare the top 15\% patients with the smallest SLD at baseline to the average patient, their risk of death is reduced by 39\% (HR=$0.61$, CI=$0.51-0.75$).

\begin{figure}[ht]
\centering
\includegraphics[scale=0.7]{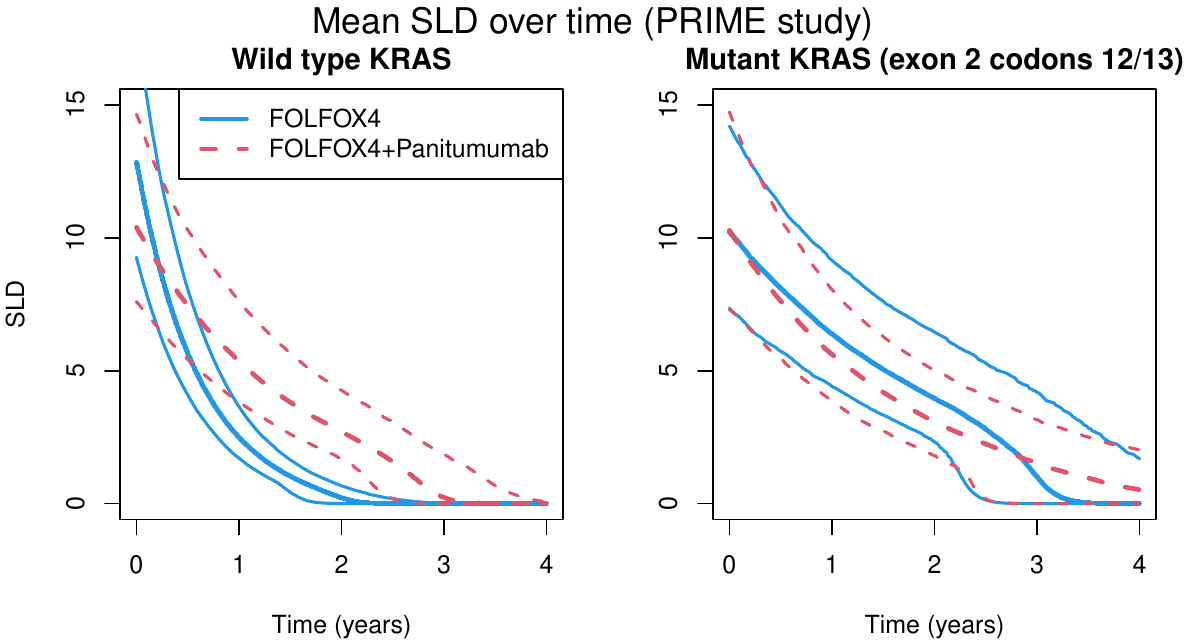}
\caption{Mean biomarker value according to treatment received for patients with wild type KRAS status (left) and mutant KRAS status (right). The 95\% credible intervals are obtained by resampling from the posterior parameter distributions.}
\label{Fig8}
\end{figure}

In conclusion, we did not find a direct effect of treatment B vs. A on the risk of death while the initial study (\citet{Douillard13}) finds a slightly significant improvement in overall survival for patients with wild type KRAS status (HR=$0.78$, CI=$0.62-0.99$), likely because of the reduced sample size available for our analysis (publicly available dataset only includes 37\% of the original set of patients). Interestingly, the analysis of the continuous part of the TPJM suggests that the reduction of the SLD over time conditional on a positive value is attenuated with treatment B compared to treatment A for patients with wild type KRAS status. A graphical representation of the mean biomarker evolution over time according to KRAS mutation status and treatment received is depicted in Figure \ref{Fig8}. It confirms the suggested significant difference between treatment arms for patients with wild type KRAS status and shows no treatment effect for patients with mutant KRAS.

\section{Discussion}

In this article, we developed a Bayesian estimation approach based on the \textbf{INLA} algorithm for two-part joint models for a longitudinal semicontinuous biomarker and a terminal event. We also provided a comparison with a frequentist alternative approach implemented into the \textbf{frailtypack} package, using small sample sizes as seen in cancer clinical trial evaluation. The frequentist estimation faced some limitations when fitting complex joint models with a high number of random effects and much increased computation time compared to \textbf{INLA}. The Bayesian estimation proposed in the R package \textbf{INLA} has been recently introduced to fit complex joint models (\citet*{vanNiekerk19b}) but to our knowledge, has never been proposed for TPJMs. Accounting for the semicontinuous nature of the biomarker, i.e. the SLD, and being able to fit joint models with more complex association structures between the biomarker and the terminal event, can be quite relevant in clinical applications by providing critical insights into the direct and indirect effect of a treatment on the event of interest. This was illustrated in our simulations and applications to two randomized cancer clinical trials.

In our simulation studies, the estimation with \textbf{INLA} was found superior to \textbf{frailtypack} in terms of computation time and precision of the fixed effects estimation. The point estimates from \textbf{frailtypack} yielded closer results to the true values of the random effects' standard deviations, the residual error term and the baseline hazard function than the posterior mean from \textbf{INLA}, even though \textbf{INLA} recovered all parameters well based on the estimated credible intervals.

Our first application to the GERCOR randomized clinical trial investigating two treatment lines to treat metastatic colorectal cancer shows some differences between the two estimation approaches. In line with our simulations, the variability of the association parameters estimates between the biomarker and the survival outcome is reduced with \textbf{INLA} compared to \textbf{frailtypack}. Moreover the computation time is reduced by a factor of more than 200 with \textbf{INLA} compared to \textbf{frailtypack} for this application to GERCOR data. The second application to the PRIME study illustrates the fact that treatment response might depend on genetic alterations or tumor biomarker status (DNA/RNA/protein features). There is now a great interest in identifying subgroups of patients with specific patterns of responses however most methods provide an average effect of covariates. Instead, our model can distinguish complete responders (i.e. SLD=0) from partial responders (i.e. SLD $>$0). This leads also to an increase in model complexity as additional covariates and random effects are included in each submodel of the TPJM. The frequentist approach proposed in \textbf{frailtypack} might have convergence issues in that situation. Interestingly, the analysis of the continuous part of the TPJM suggested that the subgroup of patients with the KRAS mutation receiving treatment B had a similar decrease of the SLD over time compared to the KRAS mutation group receiving treatment A or patients who received treatment B with wild type KRAS status. Therefore, the lack of response to the addition of anti-EGFR to FOLFOX4 chemotherapy was not fully explained by the KRAS mutation status. This could motivate further investigations of the interaction between KRAS mutation and anti-EGFR therapies to treat advanced colorectal cancer patients, in particular by including information on other somatic tumour mutations (e.g., BRAF or NRAS mutations). 

Our work has several limitations. Our applications focused on clinical trials of very advanced cancers, which often have high death rates, short follow-up and small proportions of complete responses (i.e. SLD=0). In situations where we have a higher proportion of complete responders, the relative performances of \textbf{INLA} vs. \textbf{frailtypack} could be different. The conclusions might be different for different settings (i.e. with higher zero rate and reduced censoring). For instance a meta-analysis evaluating the responses among non-Hodgkin's Lymphoma patients estimated complete response rates (i.e. SLD$=0$) ranging from $1.2\%$ to $84\%$ in the different pooled clinical trials (\citet{Mangal18}). We also notice that the two models estimated with \textbf{INLA} and \textbf{frailtypack} are not completly comparable because of the difference in the approximation of the baseline hazard function. Besides the shared random effects, other association structures have also been proposed such as the current value association, i.e., it uses the current level of the biomarker, and is available in \textbf{frailtypack}. For the TPJMs, the current value of the biomarker is defined as $\textrm{E}[Y_{ij}]=\textrm{Prob}(Y_{ij}>0)\textrm{E}[Y_{ij}|Y_{ij}>0]$, which is non linear. It cannot be directly defined as part of the latent Gaussian model and more work is warranted to include this development in \textbf{INLA}. So at the stage of development, \textbf{frailtypack} still provides more flexibility when specifying an association structure between the biomarker and the survival outcome. Besides, dynamic predictions of the event of interest is not yet implemented in \textbf{INLA} and thus requires post computations but is a major component of \textbf{frailtypack}, available for a wide range of joint models. It would be also interesting to consider a Bayesian development for the marginal TPJM we recently proposed (\citet*{Rustand21}). Finally, the definition of the hyperparameter prior distributions are an important aspect of Bayesian estimation. In this work, the PC priors provided a general setting for the priors since they provide a natural avenue to incorporate knowledge from the practitioner about the expected size of the parameter and they are constructed to be proper and avoid overfitting. Alternative prior choices for the hyperparameters can be used in \textbf{INLA} if the practitioner possesses motivation for it from expert or prior knowledge.

The reduction in the computation times with \textbf{INLA} was beyond our expectations. It improves drastically the applicability of the Bayesian estimation for complex models such as the TPJMs and other families of joint models, such as a bivariate joint model for recurrent events and a terminal event or a trivariate joint model for a longitudinal biomarker, recurrent events and a terminal event, which are currently available in \textbf{frailtypack}. Finally, \textbf{INLA} can accommodate multiple longitudinal outcomes while \textbf{frailtypack} is currently limited to a single longitudinal outcome. 

\section*{Acknowledgements}
This publication is based on research using information obtained from www.projectdatasphere.org, which is maintained by Project Data Sphere, LLC. Neither Project Data Sphere, LLC nor the owner(s) of any information from the web site have contributed to, approved or are in any way responsible for the contents of this publication. The authors acknowledge the insightful and constructive comments made by associate editor and two reviewers. These comments have greatly helped to sharpen the original submission.
\newpage
\section*{Supplementary Materials}

\begin{table}[!htb]
\centering
\textbf{Table S1: }Simulations with two correlated random effects
\scriptsize
{\tabcolsep=2.25pt
\begin{tabular}{@{}lllll@{}}
\hline
Approach &  True value &  INLA & frailtypack\\
& & Est.$^*$ (SD$^\dagger$) [CP$^{\ddagger}$] & Est. (SD) [CP]\\
\hline
\textbf{Binary part} (SLD$>$0 versus SLD=0) & & &\\
\ intercept & $\alpha_0=4$ & 4.02 (0.36) [92\%] & 4.03 (0.38) [95\%]   \\
\ time (year) & $\alpha_1=-0.5$ & -0.51 (0.12) [95\%] & -0.51 (0.12) [95\%]  \\
\ treatment (B/A) & $\alpha_2=-0.5$ & -0.50 (0.46) [95\%] & -0.50 (0.46) [95\%] \\
\ time:treatment (B/A) & $\alpha_3=0.5$  & 0.51 (0.18) [94\%] & 0.51 (0.18) [94\%]\\
\textbf{Continuous part} ($\textrm{E}[\log(Y_{ij})|Y_{ij}>0]$) & & \\
\ intercept & $\beta_0=2$ & 2.00 (0.05) [95\%] & 2.00 (0.06) [92\%] \\
\ time (years) & $\beta_1=-0.3$ & -0.30 (0.01) [95\%] & -0.30 (0.01) [94\%] \\
\ treatment (B/A) & $\beta_2=-0.3$ & -0.30 (0.08) [94\%]  & -0.30 (0.09) [91\%]\\
\ time:treatment (B/A) & $\beta_3=0.3$ & 0.30 (0.02) [95\%] & 0.30 (0.02) [94\%]\\
\ residual S.E. & $\sigma_\varepsilon=0.3$ & 0.30 (0.01) [94\%] & 0.30 (0.01) [95\%]\\
\hline
\textbf{Death risk} & & &\\
\ treatment (B/A) & $\gamma=0.2$ & 0.19 (0.28) [95\%] & 0.22 (0.36) [96\%] \\
\textbf{Association} & & \\
\ intercept (binary part) & $\varphi_a=1$ & 0.99 (0.13) [99\%] & 1.06 (0.89) [94\%] \\
\ intercept (continuous part) & $\varphi_{b_0}=1$ & 1.07 (0.15) [98\%] & 1.40 (1.28) [92\%] \\
\hline
\textbf{Random effects's standard deviation} & & &\\
\ intercept (binary part) & $\sigma_{a}=1$ & 0.93 (0.19) [92\%] & 0.96 (0.21) [94\%]\\
\ intercept (continuous part) & $\sigma_{b_0}=0.5$ & 0.51 (0.03) [96\%] & 0.50 (0.03) [87\%]\\
\ & $\rho_{ab}=0.5$ & 0.47 (0.11) [94\%] & 0.49 (0.16) [93\%]\\
\hline
\textbf{Computation time} \\
\multicolumn{2}{l}{80 CPUs (Intel Xeon Gold 6248 2.50GHz)}  & 6 sec. (1) & 74 sec. (29)\\
\textbf{\% of estimated models} & & 100\% & 91\%\\
\hline
\vspace{-0.2cm}\\
\multicolumn{4}{l}{$^*$ Posterior mean, $^\dagger$ Standard deviation of the posterior mean, $^\ddagger$ Coverage probability}
\end{tabular}}
\label{simtwo}
\end{table}

\begin{table}[!htb]
\centering
\textbf{Table S2: } Simulations with three correlated random effects
\scriptsize
{\tabcolsep=2.25pt
\begin{tabular}{@{}lllll@{}}
\hline
Approach & True value &  INLA & frailtypack\\
& & Est.$^*$ (SD$^\dagger$) [CP$^{\ddagger}$] & Est. (SD) [CP]\\
\hline
\textbf{Binary part} (SLD$>$0 versus SLD=0) & & &\\
\ intercept & $\alpha_0=4$ & 4.00 (0.36) [93\%] & 4.02 (0.41) [93\%] \\
\ time (year) & $\alpha_1=-0.5$ & -0.51 (0.12) [94\%] & -0.51 (0.13) [95\%] \\
\ treatment (B/A) & $\alpha_2=-0.5$ & -0.51 (0.48) [95\%] & -0.50 (0.51) [93\%] \\
\ time:treatment (B/A) & $\alpha_3=0.5$  & 0.51 (0.18) [95\%] & 0.51 (0.18) [95\%]\\
\textbf{Continuous part} ($\textrm{E}[\log(Y_{ij})|Y_{ij}>0]$) & & \\
\ intercept & $\beta_0=2$ & 2.00 (0.06) [95\%] & 1.99 (0.08) [85\%] \\
\ time (years) & $\beta_1=-0.3$ & -0.30 (0.06) [94\%] & -0.25 (0.12) [44\%] \\
\ treatment (B/A) & $\beta_2=-0.3$ & -0.30 (0.08) [95\%] & -0.29 (0.10) [88\%]\\
\ time:treatment (B/A) & $\beta_3=0.3$ & 0.30 (0.08) [95\%] & 0.29 (0.15) [47\%]\\
\ residual S.E. & $\sigma_\varepsilon=0.3$ & 0.30 (0.01) [93\%] & 0.30 (0.01) [95\%]\\
\hline
\textbf{Death risk} & & &\\
\ treatment (B/A) & $\gamma=0.2$ & 0.19 (0.30) [94\%] & 0.25 (0.53) [82\%]\\
\textbf{Association} & & \\
\ intercept (binary part) & $\varphi_a=1$ & 0.98 (0.12) [99\%] & 0.89 (1.86) [89\%] \\
\ intercept (continuous part) & $\varphi_{b_0}=1$ & 1.10 (0.14) [98\%] & 1.09 (1.82) [89\%] \\
\ slope (continuous part) & $\varphi_{b_1}=1$ & 1.07 (0.14) [98\%] & 1.46 (1.81) [89\%] \\
\hline
\textbf{Random effects's standard deviation} & & &\\
\ intercept (binary part) & $\sigma_{a}=1$ & 0.93 (0.15) [95\%] & 1.12 (0.29) [91\%] \\
\ intercept (continuous part) & $\sigma_{b_0}=0.5$ & 0.50 (0.03) [95\%] & 0.50 (0.04) [85\%] \\
\ slope (continuous part) & $\sigma_{b_1}=0.5$ & 0.50 (0.03) [94\%] & 0.58 (0.11) [22\%] \\
\ & $\rho_{a b_0}=0.5$ & 0.45 (0.10) [94\%] & 0.49 (0.16) [87\%] \\
\ & $\rho_{a b_1}=0.5$ & 0.44 (0.13) [93\%] & 0.57 (0.15) [73\%] \\
\ & $\rho_{b_0 b_1}=-0.2$ & -0.19 (0.10) [93\%] & -0.11 (0.19) [23\%] \\
\hline
\textbf{Computation time}\\
\multicolumn{2}{l}{80 CPUs (Intel Xeon Gold 6248 2.50GHz)} & 6 sec. (1) & 181 sec. (57)\\
\textbf{\% of estimated models} & & 100\% & 81\%\\
\hline
\vspace{-0.2cm}\\
\multicolumn{4}{l}{$^*$ Posterior mean, $^\dagger$ Standard deviation of the posterior mean, $^\ddagger$ Coverage probability}
\end{tabular}}
\label{simthree}
\end{table}

\begin{table}[!htb]
\centering
\textbf{Table S3:} Simulations with $n=500$\\
\scriptsize
{\tabcolsep=2.25pt
\begin{tabular}{@{}llllll@{}}
\hline
Approach & &  INLA & frailtypack\\
& & Est.$^*$ (SD$^\dagger$) [CP$^{\ddagger}$] & Est. (SD) [CP]\\
\hline
\textbf{Binary part} (SLD$>$0 versus SLD=0) & & &\\
\ intercept & $\alpha_0=4$ & 3.99 (0.23) [92\%] & 3.95 (0.25) [92\%] \\
\ time (year) & $\alpha_1=-0.5$ & -0.51 (0.07) [95\%] & -0.51 (0.08) [95\%] \\
\ treatment (B/A) & $\alpha_2=-0.5$ & -0.51 (0.29) [95\%] & -0.51 (0.31) [93\%] \\
\ time:treatment (B/A) & $\alpha_3=0.5$  & 0.51 (0.11) [95\%] & 0.50 (0.11) [95\%]\\
\textbf{Continuous part} ($\textrm{E}[\log(Y_{ij})|Y_{ij}>0]$) & & \\
\ intercept & $\beta_0=2$ & 2.00 (0.03) [95\%] & 1.99 (0.04) [87\%] \\
\ time (years) & $\beta_1=-0.3$ & -0.30 (0.04) [92\%] & -0.30 (0.08) [45\%] \\
\ treatment (B/A) & $\beta_2=-0.3$ & -0.30 (0.05) [96\%] & -0.30 (0.06) [89\%]\\
\ time:treatment (B/A) & $\beta_3=0.3$ & 0.30 (0.05) [95\%] & 0.29 (0.10) [46\%]\\
\ residual S.E. & $\sigma_\varepsilon=0.3$ & 0.30 (0.00) [94\%] & 0.30 (0.00) [95\%]\\
\hline
\textbf{Death risk} & & &\\
\ treatment (B/A) & $\gamma=0.2$ & 0.19 (0.18) [94\%] & 0.19 (0.28) [80\%]\\
\textbf{Association} & & \\
\ intercept (binary part) & $\varphi_a=1$ & 0.91 (0.17) [96\%] & 0.99 (1.02) [92\%] \\
\ intercept (continuous part) & $\varphi_{b_0}=1$ & 1.12 (0.18) [94\%] & 1.06 (0.96) [89\%] \\
\ slope (continuous part) & $\varphi_{b_1}=1$ & 1.09 (0.15) [97\%] & 1.08 (1.00) [89\%] \\
\hline
\textbf{Random effects's standard deviation} & & &\\
\ intercept (binary part) & $\sigma_{a}=1$ & 0.96 (0.11) [93\%] & 1.01 (0.14) [92\%] \\
\ intercept (continuous part) & $\sigma_{b_0}=0.5$ & 0.50 (0.02) [95\%] & 0.50 (0.02) [86\%] \\
\ slope (continuous part) & $\sigma_{b_1}=0.5$ & 0.50 (0.02) [95\%] & 0.53 (0.06) [29\%] \\
\ & $\rho_{a b_0}=0.5$ & 0.47 (0.08) [94\%] & 0.51 (0.11) [83\%] \\
\ & $\rho_{a b_1}=0.5$ & 0.47 (0.08) [94\%] & 0.54 (0.11) [77\%] \\
\ & $\rho_{b_0 b_1}=-0.2$ & -0.20 (0.06) [94\%] & -0.16 (0.14) [30\%]\\
\hline
\textbf{Computation time}\\
\multicolumn{2}{l}{80 CPUs (\textbf{Intel Xeon Gold 6248 2.50GHz})} & 11 sec. (1) & 340 sec. (89)\\
\textbf{\% of estimated models} & & 100\% & 96\%\\
\hline
\vspace{-0.2cm}\\
\multicolumn{4}{l}{$^*$ Posterior mean, $^\dagger$ Standard deviation of the posterior mean, $^\ddagger$ Coverage probability}
\end{tabular}}
\label{sim500}
\end{table}

\begin{table}[!htb]
\centering
\textbf{Table S4:} Simulations with splines (4 correlated random effects, 200 individuals)\\
\scriptsize
{\tabcolsep=2.25pt
\begin{tabular}{@{}lllll@{}}
\hline
Approach & &  INLA & frailtypack\\
& & Est.$^*$ (SD$^\dagger$) [CP$^{\ddagger}$] & Est. (SD) [CP]\\
\hline
\textbf{Binary part} (SLD$>$0 versus SLD=0) &\\
\ intercept & $\alpha_0=4$ & 4.08 (0.41) [94\%] & 4.12 (0.43) [96\%]\\
\ time (year) & $\alpha_1=-0.5$ & -0.51 (0.14) [94\%] & -0.52 (0.15) [94\%]\\
\ treatment (B/A) & $\alpha_2=-0.5$ & -0.51 (0.57) [93\%] & -0.53 (0.56) [95\%]\\
\ time:treatment (B/A) & $\alpha_3=0.5$  & 0.52 (0.24) [93\%] & 0.52 (0.24) [93\%]\\
\textbf{Continuous part} ($\textrm{E}[\log(Y_{ij})|Y_{ij}>0]$) & \\
\ intercept & $\beta_0=2$ & 2.00 (0.05) [95\%] & 2.00 (0.07) [91\%]\\
\ slope 1 & $\beta_1=-1$ & -0.99 (0.09) [95\%] & -1.00 (0.13) [88\%] \\
\ slope 2 & $\beta_2=-1$ & -1.00 (0.09) [92\%] & -1.02 (0.16) [85\%]\\
\ treatment (B/A) & $\beta_3=-0.3$ & -0.30 (0.06) [95\%] & -0.30 (0.08) [91\%] \\
\ slope 1:treatment (B/A) & $\beta_4=1$ & 1.00 (0.12) [96\%] & 1.00 (0.13) [94\%] \\
\ slope 2:treatment (B/A) & $\beta_5=1$ & 1.00 (0.12) [96\%] & 1.00 (0.13) [92\%]\\
\ residual S.E. & $\sigma_\varepsilon=0.3$ & 0.30 (0.01) [93\%] & 0.30 (0.01) [94\%]\\
\hline
\textbf{Death risk} &\\
\ treatment (B/A) & $\gamma=0.2$ & 0.21 (0.25) [94\%] & 0.28 (0.39) [95\%]\\
\textbf{Association} & \\
\ intercept (binary part) & $\varphi_a=0.9$ & 0.97 (0.35) [99\%] & 1.08 (1.35) [98\%]\\
\ intercept (continuous part) & $\varphi_{b_0}=1$ & 1.05 (0.39) [99\%] & 1.53 (2.37) [98\%]\\
\ slope 1 (continuous part) & $\varphi_{b_{s1}}=1$ & 1.01 (0.26) [99\%] & 1.32 (1.57) [96\%]\\
\ slope 2 (continuous part) & $\varphi_{b_{s2}}=1$ & 0.98 (0.15) [99\%] & 1.28 (2.12) [92\%]\\
\hline
\textbf{Random effects's standard deviation} & & &\\
\ intercept (binary part) & $\sigma_{a}=0.5$ & 0.55 (0.10) [98\%] & 0.78 (0.25) [89\%]\\
\ intercept (continuous part) & $\sigma_{b_0}=0.4$ & 0.42 (0.03) [93\%] & 0.40 (0.04) [89\%]\\
\ slope 1 (continuous part) & $\sigma_{b_{s1}}=0.4$ & 0.45 (0.05) [92\%] & 0.40 (0.10) [82\%]\\
\ slope 2 (continuous part) & $\sigma_{b_{s2}}=0.5$ & 0.51 (0.05) [95\%] & 0.53 (0.08) [90\%]\\
\ & $\rho_{a b_0}=0.1$ & 0.08 (0.15) [99\%] & 0.06 (0.24) [93\%]\\
\ & $\rho_{a b_{s1}}=0.1$ & 0.03 (0.17) [98\%] & 0.06 (0.29) [93\%]\\
\ & $\rho_{a b_{s2}}=0$ & -0.02 (0.20) [98\%] & -0.02 (0.28) [92\%]\\
\ & $\rho_{b_0 b_{s1}}=0.1$ & 0.04 (0.12) [95\%] & 0.16 (0.22) [85\%]\\
\ & $\rho_{b_0 b_{s2}}=0.1$ & 0.09 (0.12) [98\%] & 0.07 (0.17) [81\%]\\
\ & $\rho_{b_{s1} b_{s2}}=0.2$ & 0.16 (0.16) [97\%] & 0.22 (0.23) [83\%]\\
\hline
\textbf{Computation time}\\
\multicolumn{2}{l}{80 CPUs (\textbf{Intel Xeon Gold 6248 2.50GHz})} & 7 sec. (1) & 156 sec. (42)\\
\textbf{\% of estimated models} & & 100\% & 53\%\\
\hline
\vspace{-0.2cm}\\
\multicolumn{3}{l}{$^*$ Posterior mean, $^\dagger$ Standard deviation of the posterior mean, $^\ddagger$ Coverage probability}
\end{tabular}}
\label{simspl}
\end{table}

\begin{table}[!htb]
\centering
\textbf{Table S5: }Application of the Bayesian and frequentist two-part joint models with a linear trend to the GERCOR study with the R packages \textbf{INLA} and \textbf{frailtypack}\\
\scriptsize
{\tabcolsep=2.25pt
\begin{tabular}{@{}llllll@{}}
\hline
Approach & & INLA & frailtypack\\
& & Est.$^\dagger$ (SD$^\ddagger$) & Est. (SD)\\
\hline
\textbf{Binary part} (SLD$>$0 versus SLD=0) & & & \\
\ intercept & $\alpha_0$ & 5.21*** (0.62) & 5.10*** (0.69) \\
\ time (year) & $\alpha_1$ & -2.18*** (0.38) & -2.16*** (0.41)\\
\ treatment (B/A) & $\alpha_2$ & -1.19 (0.71) & -1.22 (0.69) \\
\ PS (1 vs. 0) & $\alpha_3$ & 2.07** (0.58) & 2.19*** (0.59)\\
\ PS (2 vs. 0) & $\alpha_4$ & 2.05 (1.11) & 1.92 (1.23) \\
\ previous\_radio (Y/N) & $\alpha_5$ & 0.92 (0.73) & 0.80 (0.71) \\
\ lung (Y/N) & $\alpha_6$ & 1.90** (0.68) & 1.83* (0.72) \\
\ time:treatment (B/A) & $\alpha_7$ & 0.34 (0.46) & 0.36 (0.47) \\
\textbf{Continuous part} ($\textrm{E}[\log(Y_{ij})|Y_{ij}>0]$) & & & \\
\ intercept & $\beta_0$ & 2.12*** (0.13) & 2.06*** (0.13) \\
\ time (years) & $\beta_1$ & -0.30*** (0.06) & -0.32*** (0.06) \\
\ treatment (B/A) & $\beta_2$ & -0.26** (0.09) & -0.27** (0.08) \\
\ PS (1 vs. 0) & $\beta_3$ & 0.38*** (0.09) & 0.37*** (0.08) \\
\ PS (2 vs. 0) & $\beta_4$ & 0.46** (0.14) & 0.44*** (0.13) \\
\ previous\_surgery (curative) & $\beta_5$ & -0.43* (0.17) & -0.38* (0.15) \\
\ previous\_surgery (palliative) & $\beta_6$ & -0.03 (0.13) & 0.04 (0.12)\\
\ previous\_radio (Y/N) & $\beta_7$ & -0.21 (0.11) & -0.16 (0.10) \\
\ metastases (metachronous vs. synchronous) & $\beta_8$ & 0.32* (0.14) & 0.30* (0.12) \\
\ time:treatment (B/A) & $\beta_9$ & 0.29*** (0.09) & 0.28*** (0.08) \\
\ residual S.E.  & $\sigma_\varepsilon$  & 0.31*** (0.01) & 0.31*** (0.01) \\
\hline
\textbf{Death risk} & & & \\
\ treatment (B/A) & $\gamma_1$ & 0.26 (0.20) & 0.23 (0.19)\\
\ PS (1 vs. 0) & $\gamma_2$ & 0.79*** (0.20) & 0.82*** (0.21) \\
\ PS (2 vs. 0) & $\gamma_3$ & 1.50*** (0.32) & 1.55*** (0.34) \\
\ previous\_surgery (curative) & $\gamma_4$& -0.89* (0.41) & -0.83* (0.41) \\
\ previous\_surgery (palliative) & $\gamma_5$ & -0.48 (0.29) & -0.44 (0.29) \\
\ metastases (metachronous vs. synchronous) & $\gamma_6$& 0.94** (0.34) & 0.96** (0.34) \\
\textbf{Association} & & & \\
\ intercept (binary part) & $\varphi_a$ & 0.12 (0.06) & 0.18 (0.14) \\
\ intercept (continuous part) & $\varphi_{b_0}$ & 0.71*** (0.16) & 0.34 (0.48) \\
\ slope (continuous part) & $\varphi_{b_1}$& 0.84*** (0.23) & 0.27 (0.72) \\
\hline
\textbf{Random effects's standard deviation} & & \\
\ intercept (binary part) & $\sigma_{a}$ & 2.78 (0.41) & 2.91 (0.40)\\
\ intercept (continuous part) & $\sigma_{b_0}$ & 0.60 (0.03) & 0.58 (0.03)\\
\ slope (continuous part) & $\sigma_{b_1}$ & 0.47 (0.06) & 0.44 (0.04)\\
\ & $\rho_{a b_0}$ & 0.47 (0.07) & 0.56 (0.07)\\
\ & $\rho_{a b_1}$ & 0.49 (0.11) & 0.51 (0.10)\\
\ & $\rho_{b_0 b_1}$ & -0.22 (0.09) & -0.19 (0.08)\\
\hline
\multicolumn{4}{l}{\textbf{Computation time (Intel Xeon Gold 6248 2.50GHz)}} \\
\ 8 CPUs & & 14 sec. & 6134 sec.\\
\ 80 CPUs & & 8 sec. & 1038 sec.\\
\hline
\vspace{-0.2cm}\\
\multicolumn{4}{l}{$^\dagger$ Posterior mean, $^\ddagger$ Posterior standard deviation}, \textsuperscript{***}$p<0.001$, 
  \textsuperscript{**}$p<0.01$, 
  \textsuperscript{*}$p<0.05$\\
\end{tabular}}
\label{resGERCORLinear}
\end{table}

\begin{figure}[!htb]
\centering
\textbf{Figure S1:} Natural cubic splines with one knot at the median of observed times for the application of the TPJM to the GERCOR study.\\
\includegraphics[scale=0.8]{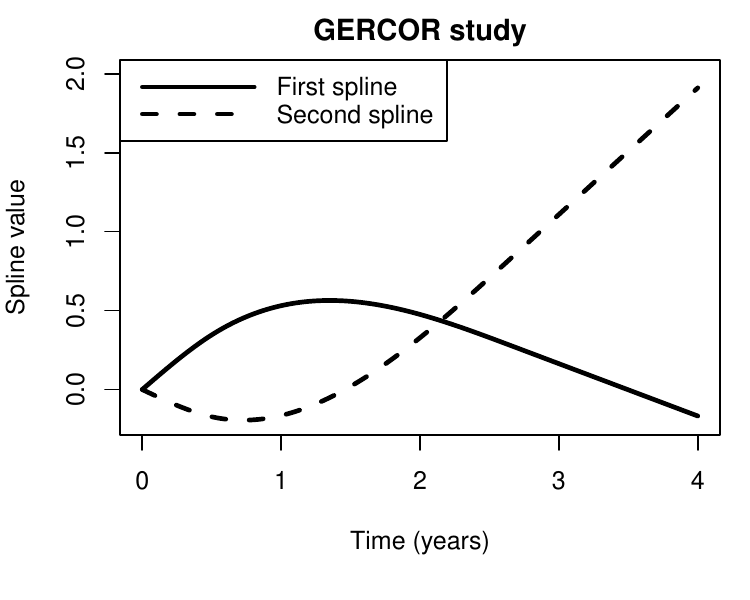}
\label{FigS1}
\end{figure}

\begin{figure}[!htb]
\centering
\textbf{Figure S2:} Observed vs. fitted longitudinal trajectories with \textbf{INLA} for five patients representative of the various observed trajectories.\\
\includegraphics[scale=0.8]{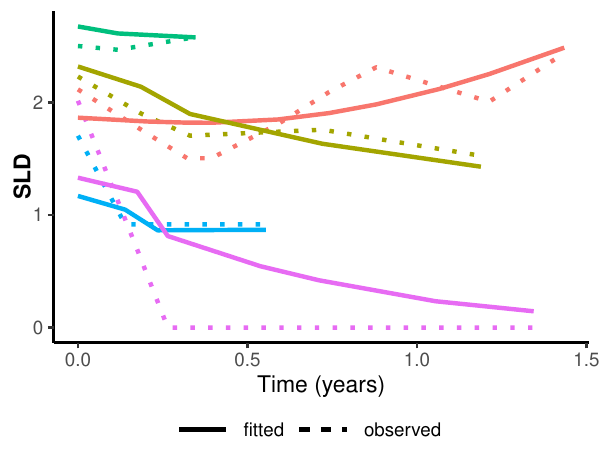}
\label{FigS2}
\end{figure}

\afterpage{\clearpage}
\newpage

\section*{R code for the estimation of the conditional TPJM with \textbf{INLAjoint} and \textbf{frailtypack}}
\begin{lstlisting}
# The code is available at github.com/DenisRustand/TPJM_sim/TPJM_INLA.R
# 1- This code shows how to simulate a dataset assuming a conditional 
#    two-part joint model with shared random effects association
# 2- The estimation of the conditional two-part joint model with INLAjoint
# 3- The estimation of the conditional two-part joint model with frailtypack
set.seed(1)

###########
###  1  ### Simulation of a dataset (scenario 3)
###########

library(mvtnorm) # for multivariate normal generation (random effects)
nsujet=200 # number of individuals
alpha_0=4 # Intercept (binary part)
alpha_1=-0.5 # slope
alpha_2=-0.5 # treatment
alpha_3=0.5 # treatment x time
beta_0=2 # Intercept (continuous part)
beta_1=-0.3 # slope
beta_2=-0.3 # treatment
beta_3=0.3 # treatment x time
sigma_e=0.3 # error term (standard error)
gamma_1=0.2 # treatmentt effect on survival
# Shared random effects association between the two-part model and survival
phi_a=1 # random intercept (binary)
phi_b=1 # random intercept (continuous)
phi_bt=1 # random slope (continuous)
baseScale=0.2 # baseline hazard scale
gap=0.4# gap between longitudinal repeated measurements
followup=4 # study duration
sigma_a=1 # random intercept (binary)
sigma_b=0.5 # random intercept (continuous)
sigma_bt=0.5 # random slope (continuous)
cor_ba=0.5 # correlation intercept (binary)/intercept (continuous)
cor_bta=0.5 # correlation intercept (binary)/slope (continuous)
cor_bbt=-0.2 # correlation continuous intercept/slope
cov_ba <- sigma_b*sigma_a*cor_ba # covariance
cov_bta <- sigma_bt*sigma_a*cor_bta
cov_bbt <- sigma_b*sigma_bt*cor_bbt
Sigma=matrix(c(sigma_a^2,cov_ba,cov_bta, # variance-covariance matrix
               cov_ba,sigma_b^2,cov_bbt,
               cov_bta,cov_bbt,sigma_bt^2),ncol=3,nrow=3)
mestime=seq(0,followup,gap) # measurement times
timej=rep(mestime, nsujet) # time column
nmesindiv=followup/gap+1 # number of individual measurements
nmesy= nmesindiv*nsujet # number of longi measurements
id<-as.factor(rep(1:nsujet, each=nmesindiv)) # patient id
# random effects generation
MVnorm <- mvtnorm::rmvnorm(nsujet, rep(0, 3), Sigma)
a_i = MVnorm[,1] # binary intercept
a_iY <- rep(a_i, each=nmesindiv) # binary intercept (repeated for longi dataset)
b_i = MVnorm[,2] # continuous intercept
b_iY <- rep(b_i, each=nmesindiv)
bt_i = MVnorm[,3] # continuous slope
bt_iY <- rep(bt_i, each=nmesindiv)
treated <- sample(1:nsujet, nsujet/2, replace=F)
trt <- as.integer(1:200 %in% sort(treated)) # treatment covariate
trtY=rep(trt, each=nmesindiv) # treatment covariate - repeated for longitudinal
## linear predictor (binary part)
linPredBin <- alpha_0+a_iY+alpha_1*timej+alpha_2*trtY+alpha_3*timej*trtY
probaBin <- exp(linPredBin)/(1+exp(linPredBin)) # probability of positive value
B <- rbinom(nmesy,1, probaBin) # observed zero values
## linear predictor (continuous part)
linPredCont <- beta_0+b_iY+(beta_1+bt_iY)*timej+beta_2*trtY+beta_3*timej*trtY
# observed biomarker values
Ypos <- rnorm(length(linPredCont), mean = linPredCont, sd = sigma_e)
Y = ifelse(B==1, Ypos, 0) # include zeros in the biomarker distribution
longDat <- data.frame(id=as.integer(id), timej, trtY, Y, B) # longitudinal dataset
## generation of exponential death times
u <- runif(nsujet) # uniform distribution for survival times generation
deathTimes <- -(log(u)/(baseScale*exp(trt*gamma_1+a_i*phi_a+b_i*phi_b+bt_i*phi_bt)))
d <- as.numeric(deathTimes<followup) # deathtimes indicator
## censoring individuals at end of follow-up (not at random)
deathTimes[deathTimes>=followup]=followup
ids <- as.factor(1:nsujet)
survDat <- data.frame(id=as.integer(ids),deathTimes, d, trt) # survival dataset
## removing longi measurements after death
ind <- rep(NA, nsujet*length(mestime))
for (i in 1:nsujet){
  for(j in 1:length(mestime)){
    if(longDat[(i-1)*length(mestime)+j, "timej"]<=survDat[i,"deathTimes"]){
      ind[(i-1)*length(mestime)+j]=1
} } }
longDat <- longDat[!is.na(ind),]
## Summary of the longitudinal and survival datasets
print(summary(survDat))
print(summary(longDat))

###########
###  2  ### Estimation of a conditional two-part joint model with INLAjoint
###########

# install instructions: https://github.com/DenisRustand/INLAjoint
library(INLAjoint)
Event <- inla.surv(survDat$deathTimes, survDat$d) # event outcome
TPinla <- joint(formLong = list(B ~ timej * trtY + (1|id),
                                Y ~ timej * trtY + (1+timej|id)),
                formSurv = Event ~ -1+trt, id = "id", timeVar = "timej",
                family = c("binomial", "gaussian"), basRisk = "rw2",
                corLong = TRUE, assoc = c("SRE_ind", "SRE_ind"),
                control=list(PriorRandom=list(r=5),
                             priorAssoc=list(mean=1, prec=10), int.strategy="eb"),
                dataLong = list(longDat, longDat[longDat$Y!=0,]), dataSurv=survDat)
summary(TPinla, sdcor=T)


###########
###  3  ### Estimation of a conditional two-part joint model with frailtypack
###########

library(frailtypack)
# kappa value (smoothing) chosen by cross-validation with an univariate Cox model
tte <- frailtyPenal(Surv(deathTimes, d)~trt, n.knots=5,
                    kappa=0, data=survDat, cross.validation=T)
kap <- round(tte$kappa,2);kap # smoothing parameter
longDat[longDat$Y==0,"Y"] <- -40 # need to set zeros as smallest value observed
# computation takes ~12min with an Intel i7-4790 (8 cores, 3.60 GHz)
TPJM <- longiPenal(formula = Surv(deathTimes, d)~trt,
                   formula.LongitudinalData = Y~timej*trtY,
                   formula.Binary=Y~timej*trtY,
                   data=survDat, data.Longi = longDat,
                   random = c("1","timej"), random.Binary=c("1"),
                   timevar="timej", id = "id",
                   link = "Random-effects", n.knots = 5, kappa = kap,
                   hazard="Splines-per", method.GH="Monte-carlo",
                   n.nodes=1000, seed.MC=1234);TPJM

\end{lstlisting}

\bibliographystyle{chicago} \bibliography{main}

\begin{thebibliography}{}

\bibitem[\protect\citeauthoryear{Andrinopoulou and Rizopoulos}{Andrinopoulou
  and Rizopoulos}{2016}]{andrinopoulou2016bayesian}
Andrinopoulou, E.-R. and D.~Rizopoulos (2016).
\newblock Bayesian shrinkage approach for a joint model of longitudinal and
  survival outcomes assuming different association structures.
\newblock {\em Statistics in medicine\/}~{\em 35\/}(26), 4813--4823.

\bibitem[\protect\citeauthoryear{Bokemeyer, Bondarenko, Hartmann, De~Braud,
  Volovat, Nippgen, Stroh, Celik, and Koralewski}{Bokemeyer
  et~al.}{2008}]{Bokemeyer08}
Bokemeyer, C., I.~Bondarenko, J.~Hartmann, F.~De~Braud, C.~Volovat, J.~Nippgen,
  C.~Stroh, I.~Celik, and P.~Koralewski (2008).
\newblock Kras status and efficacy of first-line treatment of patients with
  metastatic colorectal cancer (mcrc) with folfox with or without cetuximab:
  The opus experience.
\newblock {\em Journal of Clinical Oncology\/}~{\em 26\/}(15\_suppl),
  4000--4000.

\bibitem[\protect\citeauthoryear{Chi and Ibrahim}{Chi and
  Ibrahim}{2006}]{Chi06}
Chi, Y.-Y. and J.~G. Ibrahim (2006).
\newblock Joint models for multivariate longitudinal and multivariate survival
  data.
\newblock {\em Biometrics\/}~{\em 62\/}(2), 432--445.

\bibitem[\protect\citeauthoryear{Commenges, Joly, Gegout-Petit, and
  Liquet}{Commenges et~al.}{2007}]{Commenges07}
Commenges, D., P.~Joly, A.~Gegout-Petit, and B.~Liquet (2007).
\newblock Choice between semi-parametric estimators of markov and non-markov
  multi-state models from coarsened observations.
\newblock {\em Scandinavian Journal of Statistics\/}~{\em 34\/}(1), 33--52.

\bibitem[\protect\citeauthoryear{Douillard, Oliner, Siena, Tabernero, Burkes,
  Barugel, Humblet, Bodoky, Cunningham, Jassem, et~al.}{Douillard
  et~al.}{2013}]{Douillard13}
Douillard, J.-Y., K.~S. Oliner, S.~Siena, J.~Tabernero, R.~Burkes, M.~Barugel,
  Y.~Humblet, G.~Bodoky, D.~Cunningham, J.~Jassem, et~al. (2013).
\newblock Panitumumab--folfox4 treatment and ras mutations in colorectal
  cancer.
\newblock {\em New England Journal of Medicine\/}~{\em 369\/}(11), 1023--1034.

\bibitem[\protect\citeauthoryear{Hanson, Branscum, and Johnson}{Hanson
  et~al.}{2011}]{hanson2011predictive}
Hanson, T.~E., A.~J. Branscum, and W.~O. Johnson (2011).
\newblock Predictive comparison of joint longitudinal-survival modeling: a case
  study illustrating competing approaches.
\newblock {\em Lifetime data analysis\/}~{\em 17\/}(1), 3--28.

\bibitem[\protect\citeauthoryear{Henderson, Diggle, and Dobson}{Henderson
  et~al.}{2000}]{henderson2000joint}
Henderson, R., P.~Diggle, and A.~Dobson (2000).
\newblock Joint modelling of longitudinal measurements and event time data.
\newblock {\em Biostatistics\/}~{\em 1\/}(4), 465--480.

\bibitem[\protect\citeauthoryear{Hespanhol, Vallio, Costa, and
  Saragiotto}{Hespanhol et~al.}{2019}]{hespanhol2019understanding}
Hespanhol, L., C.~S. Vallio, L.~M. Costa, and B.~T. Saragiotto (2019).
\newblock Understanding and interpreting confidence and credible intervals
  around effect estimates.
\newblock {\em Brazilian journal of physical therapy\/}~{\em 23\/}(4),
  290--301.

\bibitem[\protect\citeauthoryear{Król, Mauguen, Mazroui, Laurent, Michiels,
  and Rondeau}{Król et~al.}{2017}]{Krol17}
Król, A., A.~Mauguen, Y.~Mazroui, A.~Laurent, S.~Michiels, and V.~Rondeau
  (2017).
\newblock Tutorial in joint modeling and prediction: A statistical software for
  correlated longitudinal outcomes, recurrent events and a terminal event.
\newblock {\em Journal of Statistical Software, Articles\/}~{\em 81\/}(3),
  1--52.

\bibitem[\protect\citeauthoryear{Król, Tournigand, Michiels, and
  Rondeau}{Król et~al.}{2018}]{Krol18}
Król, A., C.~Tournigand, S.~Michiels, and V.~Rondeau (2018).
\newblock Multivariate joint frailty model for the analysis of nonlinear tumor
  kinetics and dynamic predictions of death.
\newblock {\em Statistics in Medicine\/}~{\em 37\/}(13), 2148--2161.

\bibitem[\protect\citeauthoryear{Kurz}{Kurz}{2017}]{kurz2017tweedie}
Kurz, C.~F. (2017).
\newblock Tweedie distributions for fitting semicontinuous health care
  utilization cost data.
\newblock {\em BMC Medical Research Methodology\/}~{\em 17\/}(1), 1--8.

\bibitem[\protect\citeauthoryear{Lawrence~Gould, Boye, Crowther, Ibrahim,
  Quartey, Micallef, and Bois}{Lawrence~Gould et~al.}{2015}]{lawrence2015joint}
Lawrence~Gould, A., M.~E. Boye, M.~J. Crowther, J.~G. Ibrahim, G.~Quartey,
  S.~Micallef, and F.~Y. Bois (2015).
\newblock Joint modeling of survival and longitudinal non-survival data:
  current methods and issues. report of the dia bayesian joint modeling working
  group.
\newblock {\em Statistics in medicine\/}~{\em 34\/}(14), 2181--2195.

\bibitem[\protect\citeauthoryear{Mangal, Salem, Li, Menon, and Freise}{Mangal
  et~al.}{2018}]{Mangal18}
Mangal, N., A.~H. Salem, M.~Li, R.~Menon, and K.~J. Freise (2018).
\newblock Relationship between response rates and median progression-free
  survival in non-hodgkin's lymphoma: A meta-analysis of published clinical
  trials.
\newblock {\em Hematological Oncology\/}~{\em 36\/}(1), 37--43.

\bibitem[\protect\citeauthoryear{Marabelle, Fakih, Lopez, Shah,
  Shapira-Frommer, Nakagawa, Chung, Kindler, Lopez-Martin, Miller~Jr,
  et~al.}{Marabelle et~al.}{2020}]{Marabelle20}
Marabelle, A., M.~Fakih, J.~Lopez, M.~Shah, R.~Shapira-Frommer, K.~Nakagawa,
  H.~C. Chung, H.~L. Kindler, J.~A. Lopez-Martin, W.~H. Miller~Jr, et~al.
  (2020).
\newblock Association of tumour mutational burden with outcomes in patients
  with advanced solid tumours treated with pembrolizumab: prospective biomarker
  analysis of the multicohort, open-label, phase 2 keynote-158 study.
\newblock {\em The Lancet Oncology\/}~{\em 21\/}(10), 1353--1365.

\bibitem[\protect\citeauthoryear{Marquardt}{Marquardt}{1963}]{Marquardt63}
Marquardt, D.~W. (1963).
\newblock An algorithm for least-squares estimation of nonlinear parameters.
\newblock {\em Journal of the Society for Industrial and Applied
  Mathematics\/}~{\em 11\/}(2), 431--441.

\bibitem[\protect\citeauthoryear{Martino, Akerkar, and Rue}{Martino
  et~al.}{2011}]{martino2011approximate}
Martino, S., R.~Akerkar, and H.~Rue (2011).
\newblock Approximate bayesian inference for survival models.
\newblock {\em Scandinavian Journal of Statistics\/}~{\em 38\/}(3), 514--528.

\bibitem[\protect\citeauthoryear{Martins, Simpson, Lindgren, and Rue}{Martins
  et~al.}{2013}]{martins2013bayesian}
Martins, T.~G., D.~Simpson, F.~Lindgren, and H.~Rue (2013).
\newblock Bayesian computing with inla: new features.
\newblock {\em Computational Statistics \& Data Analysis\/}~{\em 67}, 68--83.

\bibitem[\protect\citeauthoryear{Muth, Oravecz, and Gabry}{Muth
  et~al.}{2018}]{muth2018user}
Muth, C., Z.~Oravecz, and J.~Gabry (2018).
\newblock User-friendly bayesian regression modeling: A tutorial with rstanarm
  and shinystan.
\newblock {\em Quantitative Methods for Psychology\/}~{\em 14\/}(2), 99--119.

\bibitem[\protect\citeauthoryear{Normanno, Tejpar, Morgillo, De~Luca,
  Van~Cutsem, and Ciardiello}{Normanno et~al.}{2009}]{Normanno09}
Normanno, N., S.~Tejpar, F.~Morgillo, A.~De~Luca, E.~Van~Cutsem, and
  F.~Ciardiello (2009).
\newblock Implications for kras status and egfr-targeted therapies in
  metastatic crc.
\newblock {\em Nature reviews Clinical oncology\/}~{\em 6\/}(9), 519.

\bibitem[\protect\citeauthoryear{Philipps, Hejblum, Prague, Commenges, and
  Proust-Lima}{Philipps et~al.}{2021}]{philipps2021robust}
Philipps, V., B.~P. Hejblum, M.~Prague, D.~Commenges, and C.~Proust-Lima
  (2021).
\newblock Robust and {Efficient} {Optimization} {Using} a
  {Marquardt}-{Levenberg} {Algorithm} with {R} {Package} {marqLevAlg}.
\newblock {\em The R journal\/}~{\em (accepted)}.

\bibitem[\protect\citeauthoryear{R.~Brown and G.~Ibrahim}{R.~Brown and
  G.~Ibrahim}{2003}]{r2003bayesian}
R.~Brown, E. and J.~G.~Ibrahim (2003).
\newblock A bayesian semiparametric joint hierarchical model for longitudinal
  and survival data.
\newblock {\em Biometrics\/}~{\em 59\/}(2), 221--228.

\bibitem[\protect\citeauthoryear{Rizopoulos et~al.}{Rizopoulos
  et~al.}{2016}]{rizopoulos2016r}
Rizopoulos, D. et~al. (2016).
\newblock The r package jmbayes for fitting joint models for longitudinal and
  time-to-event data using mcmc.
\newblock {\em Journal of Statistical Software\/}~{\em 72\/}(i07).

\bibitem[\protect\citeauthoryear{Rizopoulos and Ghosh}{Rizopoulos and
  Ghosh}{2011}]{Rizopoulos11}
Rizopoulos, D. and P.~Ghosh (2011).
\newblock A bayesian semiparametric multivariate joint model for multiple
  longitudinal outcomes and a time-to-event.
\newblock {\em Statistics in medicine\/}~{\em 30\/}(12), 1366--1380.

\bibitem[\protect\citeauthoryear{Rue and Held}{Rue and Held}{2005}]{Rue05}
Rue, H. and L.~Held (2005).
\newblock {\em Gaussian Markov random fields: theory and applications}.
\newblock CRC press.

\bibitem[\protect\citeauthoryear{Rue, Martino, and Chopin}{Rue
  et~al.}{2009}]{Rue09}
Rue, H., S.~Martino, and N.~Chopin (2009).
\newblock Approximate bayesian inference for latent gaussian models by using
  integrated nested laplace approximations.
\newblock {\em Journal of the Royal Statistical Society: Series B (Statistical
  Methodology)\/}~{\em 71\/}(2), 319--392.

\bibitem[\protect\citeauthoryear{Rue, Riebler, Sørbye, Illian, Simpson, and
  Lindgren}{Rue et~al.}{2017}]{Rue17}
Rue, H., A.~Riebler, S.~H. Sørbye, J.~B. Illian, D.~P. Simpson, and F.~K.
  Lindgren (2017).
\newblock Bayesian computing with inla: A review.
\newblock {\em Annual Review of Statistics and Its Application\/}~{\em 4\/}(1),
  395--421.

\bibitem[\protect\citeauthoryear{Rustand, Briollais, and Rondeau}{Rustand
  et~al.}{2021}]{Rustand21}
Rustand, D., L.~Briollais, and V.~Rondeau (2021).
\newblock {A marginal two-part joint model for a longitudinal biomarker and a
  terminal event with application to advanced head and neck cancers}.
\newblock (Under submission).

\bibitem[\protect\citeauthoryear{Rustand, Briollais, Tournigand, and
  Rondeau}{Rustand et~al.}{2020}]{Rustand20}
Rustand, D., L.~Briollais, C.~Tournigand, and V.~Rondeau (2020, 04).
\newblock {Two-part joint model for a longitudinal semicontinuous marker and a
  terminal event with application to metastatic colorectal cancer data}.
\newblock {\em Biostatistics\/}~{\em 23\/}(1), 50--68.

\bibitem[\protect\citeauthoryear{Schenk and G{\"a}rtner}{Schenk and
  G{\"a}rtner}{2004}]{Schenk04}
Schenk, O. and K.~G{\"a}rtner (2004).
\newblock Solving unsymmetric sparse systems of linear equations with pardiso.
\newblock {\em Future Generation Computer Systems\/}~{\em 20\/}(3), 475--487.

\bibitem[\protect\citeauthoryear{Simpson, Rue, Riebler, Martins, S{\o}rbye,
  et~al.}{Simpson et~al.}{2017}]{Simpson17}
Simpson, D., H.~Rue, A.~Riebler, T.~G. Martins, S.~H. S{\o}rbye, et~al. (2017).
\newblock Penalising model component complexity: A principled, practical
  approach to constructing priors.
\newblock {\em Statistical science\/}~{\em 32\/}(1), 1--28.

\bibitem[\protect\citeauthoryear{Smith, Preisser, Neelon, and
  Maciejewski}{Smith et~al.}{2014}]{Smith14}
Smith, V.~A., J.~S. Preisser, B.~Neelon, and M.~L. Maciejewski (2014).
\newblock A marginalized two-part model for semicontinuous data.
\newblock {\em Statistics in Medicine\/}~{\em 33\/}(28), 4891--4903.

\bibitem[\protect\citeauthoryear{Song, Davidian, and Tsiatis}{Song
  et~al.}{2002}]{Song02}
Song, X., M.~Davidian, and A.~A. Tsiatis (2002).
\newblock A semiparametric likelihood approach to joint modeling of
  longitudinal and time-to-event data.
\newblock {\em Biometrics\/}~{\em 58\/}(4), 742--753.

\bibitem[\protect\citeauthoryear{Spiegelhalter, Best, Carlin, and Van
  Der~Linde}{Spiegelhalter et~al.}{2002}]{Spiegelhalter02}
Spiegelhalter, D.~J., N.~G. Best, B.~P. Carlin, and A.~Van Der~Linde (2002).
\newblock Bayesian measures of model complexity and fit.
\newblock {\em Journal of the Royal Statistical Society: Series B (Statistical
  Methodology)\/}~{\em 64\/}(4), 583--639.

\bibitem[\protect\citeauthoryear{Tournigand, André, Achille, Lledo, Flesh,
  Mery-Mignard, Quinaux, Couteau, Buyse, Ganem, Landi, Colin, Louvet, and
  de~Gramont}{Tournigand et~al.}{2004}]{Tournigand04}
Tournigand, C., T.~André, E.~Achille, G.~Lledo, M.~Flesh, D.~Mery-Mignard,
  E.~Quinaux, C.~Couteau, M.~Buyse, G.~Ganem, B.~Landi, P.~Colin, C.~Louvet,
  and A.~de~Gramont (2004).
\newblock Folfiri followed by folfox6 or the reverse sequence in advanced
  colorectal cancer: A randomized gercor study.
\newblock {\em Journal of Clinical Oncology\/}~{\em 22\/}(2), 229--237.

\bibitem[\protect\citeauthoryear{Van~Cutsem, Lang, D'haens, Moiseyenko,
  Zaluski, Folprecht, Tejpar, Kisker, Stroh, and Rougier}{Van~Cutsem
  et~al.}{2008}]{VanCutsem08}
Van~Cutsem, E., I.~Lang, G.~D'haens, V.~Moiseyenko, J.~Zaluski, G.~Folprecht,
  S.~Tejpar, O.~Kisker, C.~Stroh, and P.~Rougier (2008).
\newblock Kras status and efficacy in the first-line treatment of patients with
  metastatic colorectal cancer (mcrc) treated with folfiri with or without
  cetuximab: The crystal experience.
\newblock {\em Journal of Clinical Oncology\/}~{\em 26\/}(15\_suppl), 2--2.

\bibitem[\protect\citeauthoryear{Van~Niekerk, Bakka, and Rue}{Van~Niekerk
  et~al.}{2019}]{vanNiekerk19a}
Van~Niekerk, J., H.~Bakka, and H.~Rue (2019).
\newblock Joint models as latent gaussian models-not reinventing the wheel.
\newblock {\em arXiv:1901.09365\/}.

\bibitem[\protect\citeauthoryear{Van~Niekerk, Bakka, and Rue}{Van~Niekerk
  et~al.}{2021}]{vanSankhya}
Van~Niekerk, J., H.~Bakka, and H.~Rue (2021).
\newblock Stable non-linear generalized bayesian joint models for
  survival-longitudinal data.
\newblock {\em Sankhya A\/}, 1--27.

\bibitem[\protect\citeauthoryear{Van~Niekerk, Bakka, Rue, and
  Schenk}{Van~Niekerk et~al.}{2021}]{vanNiekerk19b}
Van~Niekerk, J., H.~Bakka, H.~Rue, and O.~Schenk (2021).
\newblock New frontiers in bayesian modeling using the inla package in r.
\newblock {\em Journal of Statistical Software\/}~{\em 100}, 1--28.

\bibitem[\protect\citeauthoryear{Walker}{Walker}{1969}]{walker1969asymptotic}
Walker, A.~M. (1969).
\newblock On the asymptotic behaviour of posterior distributions.
\newblock {\em Journal of the Royal Statistical Society: Series B
  (Methodological)\/}~{\em 31\/}(1), 80--88.

\bibitem[\protect\citeauthoryear{Wulfsohn and Tsiatis}{Wulfsohn and
  Tsiatis}{1997}]{wulfsohn1997joint}
Wulfsohn, M.~S. and A.~A. Tsiatis (1997).
\newblock A joint model for survival and longitudinal data measured with error.
\newblock {\em Biometrics\/}, 330--339.

\bibitem[\protect\citeauthoryear{Xu and Zeger}{Xu and
  Zeger}{2001}]{xu2001joint}
Xu, J. and S.~L. Zeger (2001).
\newblock Joint analysis of longitudinal data comprising repeated measures and
  times to events.
\newblock {\em Journal of the Royal Statistical Society: Series C (Applied
  Statistics)\/}~{\em 50\/}(3), 375--387.

\end{thebibliography}

\end{document}